\documentclass{optica-article}

\journal{opticajournal} 

\articletype{Research Article}

\usepackage{lineno}
\usepackage{amsmath}
\usepackage{tabularx}
\usepackage{mathtools}
\usepackage{comment}
\defaultbibliography{TApaper_v0}
\defaultbibliographystyle{opticajnl}
\begin{document}

\title{Ultra-broadband transient absorption down to 200~nm enabled by soliton dynamics in gas-filled hollow capillary fibers}

\author{Pieter J. Brongers,\authormark{1} Kyle Barlow,\authormark{1} Deepjyoti Satpathy,\authormark{2} John C. Travers, \authormark{2} Christian Brahms, \authormark{2} and Malte Oppermann* \authormark{1}}

\address{\authormark{1} University of Basel, Department of Chemistry, Basel, 4056, Switzerland\\
\authormark{2} School of Engineering and Physical Sciences, Heriot-Watt University, Edinburgh, EH14 4AS, UK}

\email{\authormark{*}malte.oppermann@unibas.ch} 


\begin{bibunit} 

\begin{abstract*} 

The deep ultraviolet (DUV) window (200--300~nm) is essential for the characterization of (bio)chemical and material systems through the UV signatures of nucleobases, amino acids, peptide bonds, many organic moieties, and wide bandgap transitions. However, extending ultrafast spectroscopy to the DUV to access the associated electronic and structural dynamics has largely remained elusive, due to the limited bandwidth and efficiency of common femtosecond DUV pulse sources. We now close this gap and demonstrate ultra-broadband femtosecond transient absorption (TA) spanning 200--800~nm, achieving unprecedented coverage of the entire DUV window. We generate supercontinuum probe pulses through soliton self-compression in a helium-filled hollow capillary fiber at a repetition rate of 20 kHz and fully suppress their high intrinsic intensity fluctuations via a correlation matrix referencing scheme. We thus achieve detector-noise-limited TA measurements with an exceptional resolution of 9~\textmu OD in one second, and demonstrate these novel capabilities by resolving the ultrafast spin-crossover dynamics of a Fe(II) complex in the DUV. This work opens the path to unravel previously inaccessible photophysical and photochemical dynamics encoded in the DUV.

\end{abstract*}

\section{Introduction}

The ability to observe and control the molecular transformations that drive (bio)chemical phenomena is a defining ambition of the molecular sciences. Ultrafast spectroscopic techniques have become essential tools to achieve this on the fastest time scales, enabling the resolution of the underlying electronic and structural dynamics with femtosecond time resolution. However, although advances in laser and particle accelerator technology have enabled ultrafast measurements across the electromagnetic spectrum, from terahertz to hard X-ray, one spectral window has remained largely inaccessible: the deep ultraviolet (DUV) spanning 200--300~nm. Closing this spectral gap has been identified as a key open challenge in ultrafast spectroscopy \cite{chergui2019ultrafast,maiuri2019ultrafast}, motivated by the rich electronic and structural information encoded in characteristic UV chromophores which absorb below 300 nm: amino acids and peptide bonds in proteins, nucleobases in DNA, and many organic moieties in chemical compounds. In addition, the DUV provides access to the bandgap transitions in wide gap semiconductor materials, including several transition metal oxides, nitrides, and diamond.

A key obstacle is the limited bandwidth and efficiency of common nonlinear frequency conversion schemes in the DUV, which has restricted the generation and application of broadband femtosecond DUV pulses in ultrafast spectroscopy \cite{dubietis2017ultrafast}. Indeed, while continua in the region below 300 nm can be generated via self-phase modulation (SPM) from the second or third harmonic of Ti:sapphire lasers \cite{riedle2013electronic,borrego2018ultraviolet}, increased material dispersion and absorption strongly limit the brightness and bandwidth in the DUV. In the 250--400~nm region, higher photon fluxes can be achieved via achromatic second harmonic generation (SHG) schemes \cite{baum2004tunable,aubock2012femtosecond}, but their extension to wavelengths below 250 nm remains limited by the available SHG crystals. Additionally, third-harmonic generation (THG) in strongly confined gas cells can deliver DUV continua reaching 200 nm \cite{wanie2024ultraviolet}, but pump-probe measurements have not yet exploited the full bandwidth of this approach \cite{giovannetti2026real}. Consequently, the DUV range down to 200 nm is still a markedly underdeveloped area in ultrafast spectroscopy.

Soliton dynamics in gas-filled hollow capillary fibers (HCFs) have emerged as a promising route to address this gap by generating ultra-broadband femtosecond pulses with high brightness, excellent mode quality, and the potential to extend their bandwidth into the deep- and vacuum-UV \cite{travers2019high,travers2024optical,brahms2026ultrabroadband}. However, their adoption as DUV probe pulse sources for ultrafast spectroscopy has so far remained limited in scope: previous implementations have achieved only partial coverage of the DUV window and strong pulse-to-pulse intensity fluctuations in the continuum have severely limited their measurement sensitivities \cite{brahms2025decoupled,godinez2026ultrafast}. Consequently, soliton-driven supercontinuum sources have yet to match the noise performance of established nonlinear frequency conversion schemes in bulk media \cite{lang2018photometrics}, which have enabled advanced ultrafast spectroscopy techniques such as multi-dimensional \cite{biswas2022coherent}, chiroptical \cite{oppermann2019ultrafast}, and microscopy \cite{gross2023progress} approaches.

We now resolve this challenge through a femtosecond TA setup with a soltion-driven supercontinuum source that combines three simultaneous advances: full coverage of the DUV to near infrared (NIR) window (200-800 nm), high repetition rate (20 kHz), and detector-noise-limited measurement sensitivity. This takes soliton-driven TA to the state of the art in ultrafast spectroscopy. To this end, we employ a 20 kHz Yb:KGW laser source to drive soliton dynamics in a helium-filled stretched HCF. Through a shot-to-shot broadband detection system, we find that interference between the soliton and third-harmonic generation causes strongly correlated spectral intensity fluctuations of up to 5\% of the mean, leading to a TA measurement noise of 30 mOD ($1\sigma$) and spectral distortions of up to 0.5 mOD - an order of magnitude larger than for SPM-driven supercontinuum generation in bulk media. However, combining our single-shot detection system with a correlation matrix referencing scheme enables us to fully suppress the high-amplitude pulse-to-pulse intensity fluctuations. As a result, we achieve distortion-free TA measurements with an exceptional resolution of 9 \textmu OD in 1 second of acquisition time, which matches the highest TA measurement sensitivities reported to date \cite{lang2018photometrics}. Finally, we demonstrate these novel capabilities by measuring the ultrafast excited-state dynamics of a prototypical Fe(II)-based spin-crossover complex, which displays well-characterized TA kinetics in the DUV.

\section{Ultra-broadband transient absorption setup}

\begin{figure}[!t]
\makebox[\textwidth][c]{
\includegraphics[]{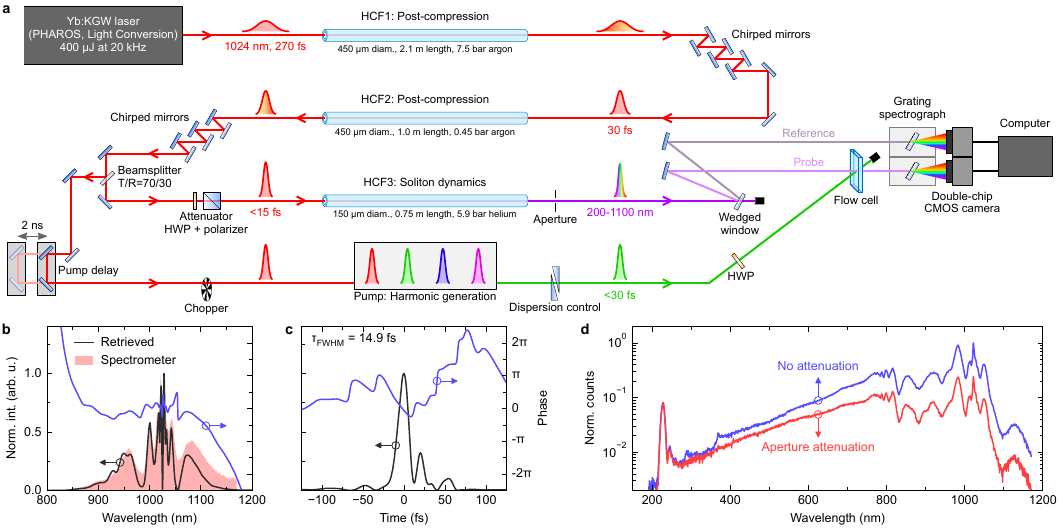}
}
\caption{Ultra-broadband transient absorption setup in the DUV. a) Schematic of the experimental setup, consisting of two post-compression HCF stages (HCF1 and HCF2), one soliton HCF stage for supercontinuum probe pulse generation (HCF3), with details provided in the text. b,c) Spectral and temporal intensity, and phase profiles of the post-compressed output of HCF2, retrieved via SHG-FROG. The phase is propagated to the entrance of HCF3. d) Typical supercontinuum spectrum obtained from HCF3, showing both an unattenuated spectrum and the effect of attenuating the output beam via an aperture with a diameter of \(\sim\)40\% of the 1/e\(^{2}\) beam diameter at 1024 nm. Due to the wavelength-dependent divergence of the output of HCF3, the NIR is attenuated more strongly than the DUV range.}
\label{fig:SetupStages_Main}
\end{figure}

The HCF setup employs a commercial diode-pumped Yb:KGW laser (LightConversion, Pharos), which delivers 400 \textmu J per pulse at 20 kHz, a center wavelength of 1024 nm, and 270 fs pulse duration. Guided by numerical simulations of pulse propagation dynamics in gas-filled HCFs via the Luna.jl software package \cite{brahmsLuna}, we designed a laser pulse source consisting of three cascaded HCF stages. Here, two post-compression HCF stages (HCF1 and HCF2) compress the NIR driver to a pulse duration of less than 15 fs, which then drives the soliton dynamics in a third HCF stage (HCF3), optimized for supercontinuum generation with high spectral uniformity down to 200 nm. The TA setup is schematically shown in Fig.~\ref{fig:SetupStages_Main}a, and Supplementary Fig.~\ref{fig:Setup06_Target_design} displays the simulated performance of each HCF from the target design. While a detailed description of the experimental setup is provided in the Supplementary Information (SI), we here focus on the most important parameters. Each HCF stage consists of a stretched fused silica capillary fiber (Molex), housed inside a home-built gas cell. The fiber dimensions, gas species, and pressure are adapted to each stage and are displayed in Fig.~\ref{fig:SetupStages_Main}a. For each HCF, the beam is focused to the optimal spot size for maximum coupling to the lowest-order transmitted mode, ensuring optimal transmission efficiency and mode quality. A piezo-controlled mirror mount (Newport, Picomotor) stabilizes the focal spot position at the fiber entrance via an imaged focal spot on a webcam and a home-built control software.

The two post-compression stages HCF1 and HCF2 spectrally broaden the NIR driver through SPM, before several dispersive mirror reflections (Ultrafast Innovations, HD120 and PC1611) negatively chirp the pulse to ensure compression at the fiber entrance of the following HCF stage. The resulting spectral and temporal profiles are characterized via a home-built SHG-FROG (see Supplementary Section~\ref{sec:SI_SetupDescription}), with the retrieved spectral and temporal intensity and phase profiles at the input of HCF3 displayed in Fig.~\ref{fig:SetupStages_Main}b,c. The excellent correspondence between the measured and retrieved data (Fig.~\ref{fig:SetupStages_Main}b and Supplementary Fig.~\ref{fig:FROG01_HCF1} and \ref{fig:FROG02_HCF2}), along with low FROG errors, demonstrate reliable retrievals. We thus determine pulse durations at full-width of half-maximum (FWHM) of 29.5 fs and 14.9 fs at the entrance of HCF2 and HCF3, respectively, in excellent agreement with the design targets (Supplementary Fig.~\ref{fig:Setup06_Target_design}). Notably, the input pulse of HCF3 (Fig.~\ref{fig:SetupStages_Main}c) displays additional satellite peaks due to third-order dispersion (TOD) accumulated between HCF2 and HCF3. However, numerical simulations (see Supplementary Section \ref{sec:SI_Comparison}) show that this has a negligible impact on the soliton dynamics driven in HCF3.

Typical input and output pulse parameters, as well as the transmission efficiencies of all HCF stages are listed in Supplementary Table \ref{tab:SetupParameters}. A cumulative throughput of approximately 70\% delivers 280 \textmu J NIR driver pulses for pump and probe pulse generation. A broadband beamsplitter (Layertec) reflects a 30\% fraction as a pump pulse, which passes an optical delay line (Aerotech) and is focused through an optical chopper (Thorlabs) to reduce its repetition rate to 10 kHz. Pump pulses at 510 nm are obtained via SHG in a 100 \textmu m thick type-I BBO and dispersion compensation is performed via a chirped mirror pair (Venteon Optics, DCM9). The pump is focused onto the sample via a concave mirror and its linear polarization is controlled through a motorized halfwave plate.

The remaining 70\% of the NIR driver is used as the probe and passes a variable attenuator, before being focused into HCF3 for supercontinuum generation. A typical output spectrum is displayed in Fig.~\ref{fig:SetupStages_Main}d (blue solid line), where the NIR driver is spectrally broadened with decreasing brightness into the DUV, before an intense resonant dispersive wave (RDW) peak is emitted near 220 nm. As the RDW's spectral position can be shifted via the gas pressure, the probe spectrum's DUV cut-off can be tuned between 190--300~nm in the present setup. To suppress the high intensity of the transmitted NIR driver, the output of HCF3 passes a variable aperture that predominantly cuts the visible to NIR range, as the output's beam divergence angle scales linearly with the wavelength \cite{nagy2008flexible}. Fig.~\ref{fig:SetupStages_Main}d shows an attenuated spectrum (red solid line) for an aperture diameter corresponding to approximately 40\% of the unclipped beam diameter in the NIR. This reduces the transmitted NIR intensity by about 75\%, while leaving the DUV mostly unperturbed. The spectral attenuation as a function of aperture size is characterized in detail in Supplementary Section \ref{sec:SI_DivergenceFocusing} and offers a straightforward approach to improve the spectral uniformity of the supercontinuum probe, particularly since suitable spectral filters remain unavailable. 

The cropped probe is reflected at near normal incidence off the front face of a wedged CaF$_{2}$ window, which reduces the total pulse energy to the nJ level required for TA spectroscopy. The reflection off the window's back face is detected as a reference for correcting shot-to-shot fluctuations in the probe. A detailed layout of the probe and reference paths is displayed in Supplementary Fig.~\ref{fig:Setup02_Referencing_layout}. Briefly, the probe is focused via an aluminum concave mirror to spot size of 50--70 \textmu m (see Supplementary Fig. \ref{fig:Setup09_Focus}) and passes through a 100 \textmu m path length flow cell (Starna) with drilled thin windows of 200 \textmu m thickness. Afterwards, the probe and reference beams are focused into short multimode fibers (Thorlabs), which are coupled to a grating-based imaging spectrograph (Chromex) equipped with a home-built double-chip CMOS detector for simultaneous probe and reference detection at 20 kHz. To avoid overlaps with higher diffraction orders, probe and reference spectra are recorded in two separate spectral regions with different gratings: 1) the DUV region spanning 200--400~nm, and 2) the visible region spanning 400--800~nm. Typical recorded probe and reference spectra are displayed in Fig.~\ref{fig:Noise_Probe_Main}a. A uniform probe spectrum is achieved in the DUV, as the cumulated spectral efficiencies of the employed mirrors, multimode fibers, gratings, and CMOS chips reduce the intensity of the RDW compared to the 230--400~nm range. In the visible region, the spectral contrast is higher, as both the probe's spectral intensity and the detector's efficiency increase substantially from 400 nm to 800 nm. The setup thus delivers full coverage of the DUV window down to 200 nm with high brightness, spectral uniformity, and shot-to-shot data acquisition at 20 kHz - pre-requisites for attaining high measurement sensitivities in TA.

\section{Characterization and suppression of supercontinuum fluctuations} \label{sec:Main_Noise}

\begin{figure}[!t]
\centering
\makebox[\textwidth][c]{
\includegraphics[]{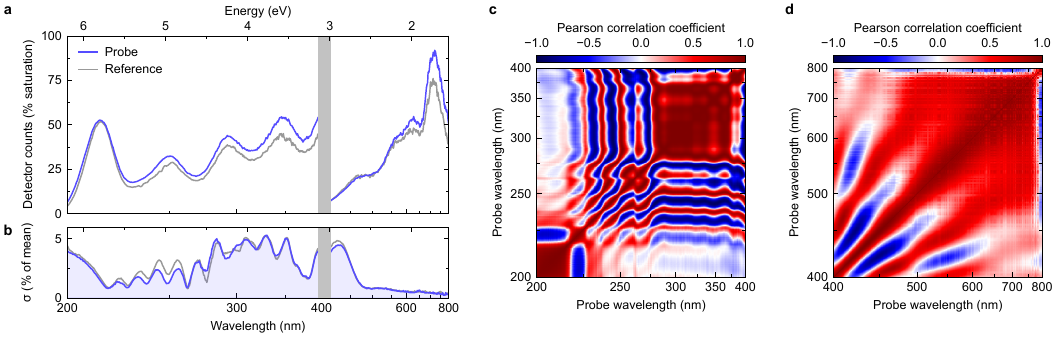}
}
\caption{Spectrally-resolved probe light statistics derived from 30'000 individual laser shots. The 200--400~nm and 400--800~nm wavelength regions have been detected in separate measurements and a gray bar is used to separate the data. a) Mean of the detected intensity of the probe and reference. b) Standard deviation of the probe and reference intensity expressed as a percentage of the mean. c,d) Probe-probe correlation matrices showing the Pearson correlation coefficient in the DUV (c) and visible (d) spectral region.}
\label{fig:Noise_Probe_Main}
\end{figure}

In TA spectroscopy, the achievable measurement sensitivity critically depends on the shot-to-shot spectral intensity fluctuations of the probe pulses \cite{dobryakov2010femtosecond,bradler2014temporal,brazard2015accurate,lang2018photometrics}: the standard deviation of the TA signal is directly proportional to that of the detected probe intensities, whereas strongly correlated intensity fluctuations between different probe wavelengths (spectral correlations) can distort TA spectra and thus cause measurement artifacts. For soliton-driven supercontinuum generation, numerical simulations indicate that THG from the self-compressed NIR driver leads to high-amplitude spectral energy density fluctuations up to the 10\% scale, which can interfere with adjacent spectral regions \cite{brahms2021timing}. Indeed, previous studies using a soliton-driven supercontinuum displayed pronounced spectral distortions of TA spectra in the THG region of the driver \cite{brahms2025decoupled,godinez2026ultrafast}, suggesting that strong spectral correlations may limit the achievable measurement sensitivity. However, the associated shot-to-shot spectral intensity fluctuations of HCF-based supercontinuum sources have not yet been fully characterized.

We perform a complete noise analysis of our TA setup, determine the spectral correlations in the supercontinuum probe, and assess the efficiency of our referenced detection system to suppress the resulting TA measurement noise and spectral distortions. The shot-to-shot total energy noise of the NIR driver increases only marginally through the compression stages, from 0.17\% at the laser source to 0.22\% after HCF2 (see Supplementary Fig.~\ref{fig:Noise_SI_01_stats_fund}). The mean intensity and standard deviation of 30'000 consecutive supercontinuum shots generated in HCF3 are presented in Fig.~\ref{fig:Noise_Probe_Main}a,b. A complete statistical analysis at selected probe wavelengths is displayed in Supplementary Fig.~\ref{fig:Noise_SI_Intensity_stats}. The observed noise signatures reveal three distinct spectral regions, which can be associated with the three major nonlinear processes contributing to the soliton-generated supercontinuum \cite{travers2019high,travers2024optical,brahms2026ultrabroadband}. 

Region 1 (480--800~nm) is dominated by SPM of the soliton and exhibits low-amplitude noise <1\%, consistent with SPM-based continuum generation in condensed-phase materials \cite{keller2023high}. Region 2 (230--480~nm) is associated with THG from the self-compressed NIR driver and shows strongly modulated noise amplitudes reaching up to 5\%, in qualitative agreement with theoretical predictions \cite{brahms2021timing}. Notably, the strongly modulated noise signature of the THG covers a much broader spectral region than might be expected from the initial bandwidth of the NIR driver (Fig.~\ref{fig:SetupStages_Main}b). Our simulations (Supplementary Fig. \ref{fig:Setup07_Simulations_HCF3}a) show that as the NIR driver broadens along the HCF, so does the bandwidth of its third harmonic, reaching approximately 160--500 nm at the fiber exit. In the extreme case at the self-compression point, the broadened soliton thus generates an intense, phase-locked and octave-spanning third harmonic, which interferes with the soliton supercontinuum \cite{savitsky2023sub}. As a consequence, the spectral intensity in the overlap region scales with a $f-3f$ interference term $\cos{(\phi_\textrm{sol}(\omega)-\phi_\textrm{THG}(\omega)-2\phi_\textrm{CEP})}$, where $\phi_\textrm{sol}$ and $\phi_\textrm{THG}$ denote the spectral phases of the soliton driver and the generated THG, and $\phi_\textrm{CEP}$ the driver’s carrier envelope phase (CEP). Since the laser employed in our setup is not CEP-stabilized, this interference directly maps the random CEP to spectral intensity noise. This is supported by the spectrally-resolved probe intensity distributions in Supplementary Figs. \ref{fig:Noise_SI_Intensity_stats} and \ref{fig:Noise_SI_all_histograms}, which show a bimodal distribution in the THG region, in agreement with the expected result from randomly sampling a sinusoidal function (see Supplementary Fig. \ref{fig:Noise_SI_cos_Bimodal}).

Finally, region 3 (200--230~nm) is dominated by RDW emission and displays a moderate noise of $\approx$~1.5\% at the RDW peak which then increases to about 4~\% at the high-energy cut-off at 200~nm. This noise enhancement has been attributed to variations in the position of the self-compression point caused by energy fluctuations of the NIR driver \cite{brahms2021timing}: if the self-compression point is near the HCF exit, as in HCF3, small variations in its position can modulate the total energy transferred to the RDW band, which is most efficient at the self-compression point.

To characterize the spectral correlations in the soliton supercontinuum fluctuations, Fig. \ref{fig:Noise_Probe_Main}c,d displays the Pearson correlation coefficient between signals measured on different probe detector pixels (see also Supplementary Figs. \ref{fig:Noise_SI_02_corr_probe} and \ref{fig:Noise_SI_03_corr_shots}). The SPM-dominated region 1 mainly displays positive correlations, whose moderate off-diagonal values are consistent with previous investigations of SPM-driven supercontinua \cite{bradler2014temporal}. The THG-domintated region 2 shows a complex correlation structure with very high positive correlation values across its central 280--370 nm range and pronounced correlation modulations in its outer high- and low-energy edge regions. The high self-correlation in the central THG region can likely be attributed to a wavelength-independent CEP dependence of the associated $f-3f$ interference term \cite{savitsky2023sub}. However, the origin of the adjacent correlation modulations remains unclear and its elucidation is beyond the scope of the present study. Finally, the spectral correlations in the RDW-dominated region 3 are characterized by a sign-change at its low-energy edge. This can be attributed to spectral shifts in the RDW emission caused by pulse-to-pulse energy fluctuations in the NIR driver. Notably, the RDW displays a near-zero correlation with the THG region, suggesting that their intensity fluctuations are largely decoupled. This is consistent with the earlier observations that noise in the RDW is mainly caused by pulse energy fluctuations in the NIR driver, whereas noise in THG is largely due to the random CEP fluctuations.

\begin{figure}[!t]
\centering\includegraphics{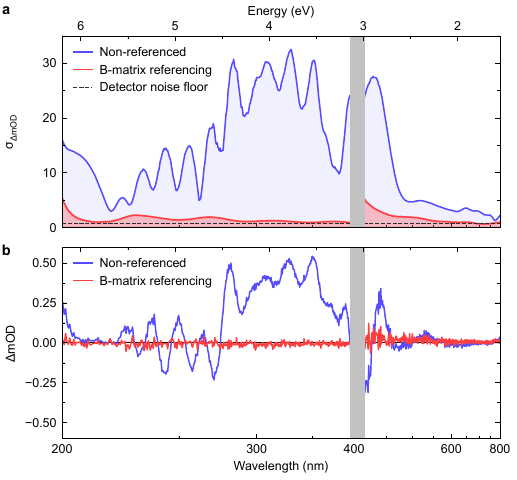}
\caption{Spectrally-resolved TA statistics derived from the same 30'000 individual laser shots as in Fig. \ref{fig:Noise_Probe_Main}. The 200--400~nm and 400--800~nm wavelength regions have been detected in separate measurements and a gray bar is used to separate the data. a) TA baseline measurement without referencing and with B-matrix referencing. b) TA standard deviation without referencing and with B-matrix referencing. The dashed line at \(\sigma_{\Delta\text{mOD}} =\) 0.8 indicates the detector noise floor for 100\% detector saturation for non-referenced data.}
\label{fig:Noise_TA_Main}
\end{figure}

We thus find that for probe wavelengths below 500 nm, the noise of the soliton supercontinuum is up to an order of magnitude larger than for continuum sources driven purely by SPM, while THG causes pronounced spectral correlations in regions where it interferes with the soliton. To quantify the impact of these noise characteristics on TA measurements, Fig. \ref{fig:Noise_TA_Main} plots the TA standard deviation $\sigma_{\Delta\text{mOD}}$ and the mean TA baseline in the absence of a pump, calculated from the series of probe shots analyzed in Fig. \ref{fig:Noise_Probe_Main}a,b (blue solid lines). A complete statistical analysis at selected probe wavelengths is displayed in Supplementary Figs.~\ref{fig:Noise_SI_04_stats_215}--\ref{fig:Noise_SI_04_stats_650}. As expected, $\sigma_{\Delta\text{mOD}}$ closely follows the standard deviation of the probe (see Supplementary Eq. \ref{eqn:Noise_nonRefODNoise}), reaching up to 30~mOD. Notably, the TA baseline displays strongly modulated pseudo-structures with amplitudes of up to 0.5~mOD in the THG region, in agreement with its spectral correlations and high-amplitude noise. In comparison, the TA pseudo-structures below 230 nm and above 500 nm are nearly negligible, due to reduced spectral correlations and intensity noise.

To suppress the TA noise and baseline distortions, we implemented a correlation matrix referencing scheme (also called \textit{B-matrix} referencing), which is known to be particularly efficient in correcting high-amplitude supercontinuum fluctuations with complex correlation structures \cite{feng2017general,feng2019optimized}. The complete spectral covariance structure between a series of single-shot reference and probe spectra is measured to calculate the so-called B-matrix, which optimally weights all reference pixels to correct intensity noise for each probe pixel. In this way, B-matrix referencing can reach the single-detector noise floor and efficiently remove pseudo-structures from TA measurements (for further details see section~\ref{sec:SI_Noise} in the SI). This is demonstrated in Fig. \ref{fig:Noise_TA_Main}, which displays the standard deviation and TA baseline obtained with B-matrix referencing (solid red line), along with the detector noise floor (dashed black line) at 0.8~mOD (see Supplementary Section \ref{sec:SI_Noise_detector}). We thus achieve detector-noise-limited, artifact-free TA measurements with an average standard deviation around 0.9~mOD. In 1~second of acquisition time at 20~kHz, the setup therefore provides a standard error of 9~\textmu OD, which matches the current state of the art in TA spectroscopy \cite{lang2018photometrics}.

\section{Ultra-broadband transient absorption}

\begin{figure}[t!]
\centering
\makebox[\textwidth][c]{
\includegraphics[]{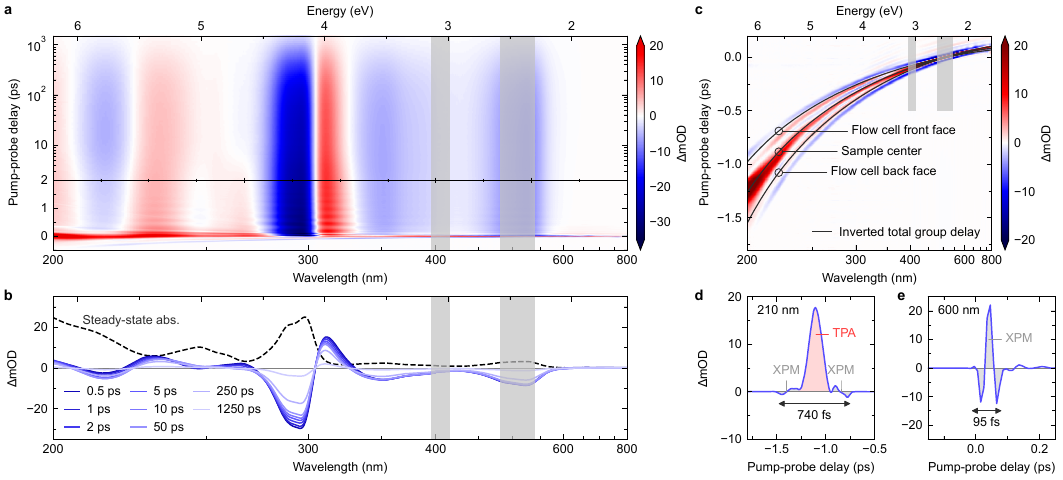}
}
\caption{Transient absorption measurements in the DUV. a) TA signal from [Fe(bpy)\(_{3}\)]Cl\(_{2}\) in H\(_{2}\)O, photoexcited at 510 nm and with time zero correction. b) Transient spectra at selected pump-probe delays taken as horizontal slices of the map in (a). The dotted line indicates the steady-state absorption spectrum with arbitrary scaling. (c) TA signal from pure H\(_{2}\)O prior to time zero correction. Plotted on top of the map is the accumulated group delay of the probe between the exit of HCF3 and the front, center and back of the pump-probe interaction volume, shown relative to the value at 510 nm. (d-e) Traces of (c) at two selected wavelengths, illustrating the contributions from two-photon absorption (TPA) and cross-phase modulation (XPM). The temporal width of the coherent artifact is indicated. The shaded areas in (a-c) at 400 nm and 510 nm indicate where two separate measurements are merged and pump scatter may occur, respectively.}
\label{fig:TA_Main}
\end{figure}

To demonstrate the capabilities of the TA setup, we performed proof-of-concept measurements on the prototypical spin-crossover complex [Fe(bpy)\(_{3}\)]Cl\(_{2}\) in H\(_{2}\)O, due to its well-characterized excited-state dynamics with prominent TA features in the DUV region \cite{gawelda2007ultrafast,smeigh2008femtosecond,consani2009vibrational,aubock2015sub,gawelda2007structural,bressler2009femtosecond,huse2011femtosecond,zhang2014tracking,lemke2017coherent,oppermann2022chiral}. Here, photo-excitation of the singlet metal-to-ligand charge-transfer (MLCT) band near 520 nm leads to an ultrafast spin-crossover to the metal-centered high-spin (HS; spin angular momentum S~=~2) state in less than 50 fs with near unity efficiency. The HS state then relaxes back to the low-spin (S~=~0) ground state on the subnanosecond scale in aqueous solution. Previous TA measurements have established that photo-excitation to the HS state redshifts the absorption bands of the ligand sphere, serving as a spectroscopic marker in the DUV that tracks its electronic and vibrational dynamics \cite{consani2009vibrational,aubock2015sub,oppermann2022chiral}.

All relevant details on the sample preparation, measurement conditions, and data processing are described in Supplementary Section~\ref{sec:SI_Ultrabroadband_Transient_absorption}. Fig.~\ref{fig:TA_Main}a displays the TA map obtained under photo-excitation at 510 nm with B-matrix referencing and after correction of the probe's group velocity dispersion (GVD). Fig.~\ref{fig:TA_Main}b shows TA spectra at selected pump-probe delays, highlighting the excellent signal-to-noise of the measurement. We performed a global lifetime analysis (GLA) on the data (see Supplementary Section \ref{sec:SI_GLA}), with the obtained decay-associated and species-associated difference spectra (DADS and SADS) and the fit residuals shown in Supplementary Fig.~\ref{fig:TA02_GLA_fits}. Note that the fits do not include the coherent oscillations observed across the DUV window, which have previously been assigned to the impulsively excited HS state \cite{consani2009vibrational}. A minimum of three sequential exponential decays are required to accurately fit the data, giving the time constants \(\tau_{1}\) = 0.65\(\pm\)0.2~ps, \(\tau_{2}\) = 3.67\(\pm\)0.3~ps, and \(\tau_{3}\) = 664\(\pm\)5~ps. These values are in excellent agreement with previous TA measurements and correspond to vibrational cooling within the HS state (\(\tau_{1}\), \(\tau_{2}\)) and HS-state decay to the ground state (\(\tau_{3}\)) \cite{consani2009vibrational,aubock2015sub,oppermann2022chiral}. Notably, the HS-state kinetics extend over the entire DUV region. This suggests that the entire DUV window is dominated by the red-shift of the ligand-centered transitions of the complex in the HS state, consistent with the strong ground-state absorption of the ligands below 300 nm.

To assess the time resolution of the experiments, we recorded the TA of neat H\(_{2}\)O under the same conditions. The data is displayed in Fig. \ref{fig:TA_Main}c without GVD correction and shows the coherent artifact (CA) arising from nonlinear pump--probe interactions in the sample flow cell during their temporal overlap \cite{beckwith2020data}. The full temporal width of the CA remains limited to \(\sim\)100 fs in the 490--800 nm region, below 400 fs for 315--490 nm, and increases to a maximum of 860 fs at 200 nm (see Fig. \ref{fig:TA_Main}c and Supplementary Fig. \ref{fig:IRF01_Experimental_time_resolution}). The practical time resolution of the setup corresponds to half of the temporal width of the CA, the first pump-probe delay where a CA-free TA signal can be observed. Below 500 nm, multiple cross-phase modulation (XPM) features can be identified, while below 250 nm a strong positive two-photon absorption (TPA) feature emerges at the center of the CA (see Fig. \ref{fig:TA_Main}c). To rationalize the strongly increased width of the CA in the DUV, we performed a detailed chirp analysis of the probe continuum (see Supplementary Section \ref{sec:SI_IRF}). As shown in Fig. \ref{fig:TA_Main}c the accumulated group delay at the sample center matches the center of the CA, while the group delay difference between the front and back faces of the flow cell scales with the width of the CA. We thus suggest that in the DUV, the width of the CA is dominated by the walk-off between the pump and probe pulses along the interaction path, which temporally separates the XPM contributions from the front and back faces of the flow cell \cite{ekvall2000cross}. Consequently, the total width of the CA becomes larger than the cross-correlation of the pulses, represented by the width of the central TPA feature (see Fig. \ref{fig:TA_Main}c,d). This indicates that the width of the CA overestimates the setup's effective response function (ERF) \cite{polli2010effective,liebel2015principles}, which determines the fastest rate of change in a TA signal that can be resolved. As upper limits to the ERF in the DUV, we find that Gaussian fits to the TPA feature at 200 nm and 210 nm result in 160 fs and 150 fs at FWHM, respectively, while coherent oscillations from SiO\textsubscript{2} with a 55 fs period are resolved in the 300--400 nm range. To further improve the time resolution of the setup, the pump-probe walk-off can be reduced through shorter sample path lengths and liquid sample jets without windows \cite{godinez2026ultrafast}, and the overall dispersion of the probe can be reduced by running the experiment in a vacuum noble-gas-purged environment \cite{brahms2025decoupled}.

\section{Conclusion}

We have demonstrated ultra-broadband transient absorption enabled by soliton dynamics in hollow capillary fibers in the range 200--800 nm, thereby achieving an unprecedented coverage of the entire DUV window (200--300 nm). The setup delivers femtosecond time resolution and detector-noise-limited TA measurement sensitivity at a 20 kHz repetition rate, reaching the current state of the art in ultrafast spectroscopy. Through a statistical analysis of the pulse-to-pulse spectral intensity fluctuations of the soliton supercontinuum, we revealed that highly correlated noise structures cause TA noise of up to 30 mOD ($1\sigma$) with spectral baseline distortions of up to 0.5 mOD - an order of magnitude larger than for established probe sources driven by self-phase modulation in bulk media. Although this has previously restricted the resolution of soliton-based ultrafast spectroscopy setups, we overcame these limitations through a correlation matrix referencing scheme, which measures and suppresses the complex noise structure intrinsic to the soliton dynamics. On this basis, we achieved artifact-free TA measurements with an exceptional resolution of 9~\textmu OD in 1 second of acquisition time and demonstrated these novel capabilities by measuring the ultrafast excited-state dynamics of a prototypical Fe(II)-based spin-crossover complex via its spectral fingerprints in the DUV.

In conclusion, this work unlocks the full potential of soliton-driven probe pulse generation for ultrafast spectroscopy, by extending its broadband spectral range to 200 nm for the first time, and by achieving multi-kHz shot-to-shot detection at the detector noise limit. These advances open the path to extend advanced ultrafast and nonlinear spectroscopic techniques to the DUV range, and to resolve previously inaccessible electronic and structural dynamics encoded in important UV chromophores, including DNA nucleobases, amino acids, peptide bonds, and a wide range of organic ligands, as well as in the bandgap transitions in wide gap semiconductor materials.

\clearpage

\begin{backmatter}
\bmsection{Funding}

Swiss National Science Foundation (215156 and 229589). The Royal Society (IES$\backslash $R3$\backslash$223054). The University of Basel and its Department of Chemistry. The Royal Academy (Research Fellowship RF/202122/21/133). The European Research Council (Starting Grant agreement FASTER, 101161675).

\bmsection{Acknowledgments}

We thank Michael Zengaffinen from the University of Basel for the preparation of the [Fe(bpy)\(_{3}\)]Cl\(_{2}\) complex.

\bmsection{Disclosures}

The authors declare no conflicts of interests.

\bmsection{Data Availability Statement}

Data underlying the results presented in this paper are not publicly available at this time but may be obtained from the authors upon reasonable request.

\bmsection{Supplemental document} See supplementary information for supporting content. 

\end{backmatter}



\putbib

\end{bibunit}

\clearpage

\setcounter{page}{1}
\newpage

\setcounter{subsection}{0} \renewcommand{\thesubsection}{S.\arabic{subsection}}
\setcounter{figure}{0} \renewcommand{\thefigure}{S.\arabic{figure}}
\setcounter{table}{0} \renewcommand{\thetable}{S.\arabic{table}}
\setcounter{equation}{0} \renewcommand{\theequation}{S.\arabic{equation}}

\begin{bibunit}

\section*{\large Supplementary information for "Ultra-broadband transient absorption down to 200~nm enabled by soliton dynamics in gas-filled hollow capillary fibers"}

\author{Pieter J. Brongers,\authormark{1} Kyle Barlow,\authormark{1} Deepjyoti Satpathy,\authormark{2} John C. Travers, \authormark{2} Christian Brahms, \authormark{2} and Malte Oppermann* \authormark{1}}

\address{\authormark{1} University of Basel, Department of Chemistry, Basel, 4056, Switzerland\\
\authormark{2} School of Engineering and Physical Sciences, Heriot-Watt University, Edinburgh, EH14 4AS, UK}

\email{\authormark{*}malte.oppermann@unibas.ch}

\etocsettocdepth{subsection}
\localtableofcontents

\subsection{Design and implementation of transient absorption setup} 

\subsubsection{Initial design of HCF stages} \label{sec:SI_Simulations_HCFs}

Soliton dynamics are driven by a combination of self-phase modulation and anomalous dispersion at the driver wavelength. Higher-order solitons spectrally broaden and temporally self-compress up to a point of maximum self-compression, where the soliton breaks up through the emission of a resonant dispersive wave (RDW). At this point of soliton fission, the spectrum exhibits the largest bandwidth, spanning from the driving pump to the RDW, making the peak position of the RDW emission (\(\lambda_{\text{RDW}}\)) a measure of the short-wave cut-off of the supercontinuum. The propagation distance from the fiber entrance to the fission point is quantified by the fission length \(L_{\text{fiss}}\). To extract the maximum bandwidth of the self-compressed supercontinuum pulse, \(L_{\text{fiss}}\) should be close to the total length of the fiber. The spectral position of the RDW depends on higher-order linear and nonlinear dispersion \cite{erkintalo2012cascaded,dudley2006supercontinuum} and generally shifts to shorter wavelengths by reducing the gas pressure \cite{travers2019high}. However, the reduced gas pressure lowers the degree of nonlinearity, which can only be compensated by a higher peak intensity up to the ionization limit of the gas. Instead, to reach the soliton self-compression regime in gas-filled hollow capillary fibers (HCFs) with practical dimensions, the input pulse duration has to decrease, typically to below 15 fs \cite{travers2024optical}. More detailed accounts of the underlying nonlinear optical processes and the resulting practical considerations and scaling rules can be found in recent publications and references therein \cite{travers2019high,travers2024optical,brahms2022efficient}.

\begin{figure}[t!]
\centering\includegraphics[]{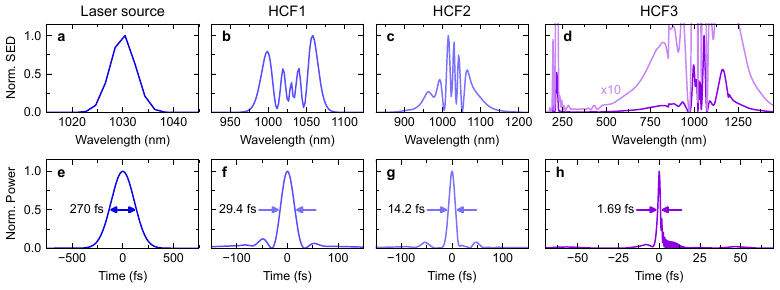}
\caption{Simulations of the design for two post-compression stages (HCF1 and HCF2) and one soliton stage (HCF3). Normalized spectral energy density (SED) (a-d) and temporal profile (e-h) for the laser output and each HCF. The pulses of HCF1 and HCF2 in (f,g) are dispersion compensated and propagated to the entrance of the next fiber. The pulse of HCF3 is at the exit of the fiber. The spectrum of the laser output is jagged due to the sampling density.}
\label{fig:Setup06_Target_design}
\end{figure}

Nonlinear optical dynamics simulations in the Luna.jl software package \cite{brahmsLuna} are performed to determine suitable design parameters for gas-filled hollow core fiber stages with the aim to first compress an NIR driver pump to below 15 fs and then generate a supercontinuum probe through soliton self-compression. As the starting point, the output of the employed driving laser source is approximated as a Fourier-Transform-limited (FTL) Gaussian pulse, centered at 1030 nm, with a pulse duration of 270 fs, derived from autocorrelation measurements provided by the supplier. To compress the NIR pump, we choose to employ a cascade of two HCF-based post-compression stages, such that each stage's compression factor remains below 10 and operates in a regime dominated by self-phase modulation. For the supercontinuum generation, the primary target is to achieve a spectral extension to 200 nm with highest possible spectral uniformity without discontinuities in a 200--800~nm probe window. In this respect, it was found that favorable soliton dynamics can be obtained with NIR driver pulses of approximately 15 fs duration at input energies less than 100 \textmu J. On this basis, we determined suitable HCF parameters to achieve the above design criteria, including the HCF dimensions, and the gas species and pressure. Table \ref{tab:SetupParameters} list the design specifications, whereas Fig. \ref{fig:Setup06_Target_design} shows the resulting simulated spectra and pulse durations of the outputs of the laser source, the two consecutive post-compression stages (HCF1 and HCF2), and the soliton stage (HCF3).

\subsubsection{Implementation of the HCF-based transient absorption setup} \label{sec:SI_SetupDescription}

\begin{table}[htbp]
\centering
\begin{threeparttable}
\caption{HCF specifications and performance.}
{\small 
\begin{tabular}{cccc}
\hline
 & HCF1 & HCF2 & HCF3 \\
\hline
\multicolumn{4}{c}{Capillary parameters} \\
 \hline
Model & Molex, TSP450670 & Molex, TSP450670 & Molex, TSP150375 \\
Diameter (\textmu m) & 450 & 450 & 150 \\
Length (m) & 2.1 & 1.0 & 0.75 \\
Gas & Argon & Argon & Helium \\
Pressure (bar), setup design & 7.0 & 0.50 & 7.0 \\
Pressure (bar), experiment & 7.5 & 0.45 & 5.9 \\
\hline
\multicolumn{4}{c}{Bandwidth and pulse duration} \\
\hline
Bandwidth (nm) & 950--1100 & 850--1200 & 200--1200<\tnote{a} \\
Pulse duration (fs) & 29.5 & 14.9 & \textit{Not characterized} \\
\hline
\multicolumn{4}{c}{Transmission performance} \\
\hline
Typical output energy (\textmu J) & 330\tnote{b} & 280\tnote{b} & 45\tnote{c} \\
Typical transmission & 86\% & 86\% & 36\%\tnote{d} \\
HE\(_{11}\) transmission at 1024 nm & 92\% & 96\% & 45\% \\
Maximum CE\tnote{e} & 93\% & 90\% & 80\% \\
\hline
\end{tabular}
}
\begin{tablenotes}\footnotesize{
    \item[a] Determining the extend of the NIR is limited by the spectrometer's detector range.
    \item[b] At operating pressure.
    \item[c] At vacuum and corrected for reflections of the output window.
    \item[d] Corrected for reflections of the output window.
    \item[e] Coupling efficiency assuming solely launching into and transmission of the HE\(_{11}\) mode.}
\end{tablenotes}
\label{tab:SetupParameters}
\end{threeparttable}
\end{table}

\begin{figure}[t!]
\centering\includegraphics[width=\textwidth]{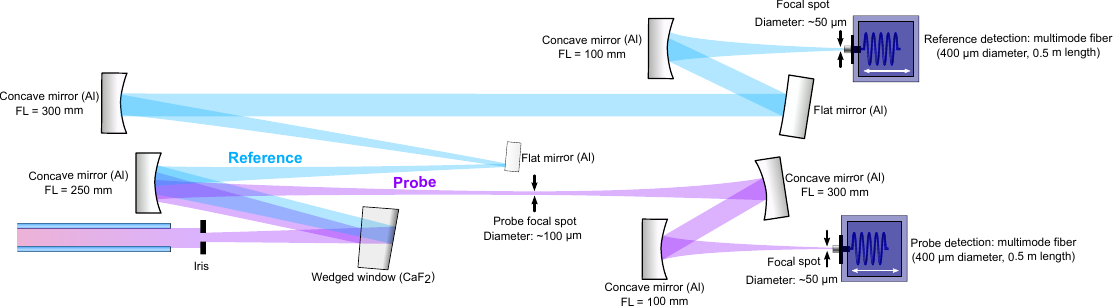}
\caption{Detailed layout of the probe and reference line following HCF3. Note that the gas cell and cell windows are not displayed for simplicity.}
\label{fig:Setup02_Referencing_layout}
\end{figure}

Figs. \ref{fig:SetupStages_Main}a and \ref{fig:Setup02_Referencing_layout} display a schematic of the layout of the HCF stages and the TA detection scheme, respectively. Table \ref{tab:SetupParameters} lists HCF parameters and typical performance measures. In brief, a cascade of two HCF post-compression stages are used to obtain \(\sim\)15 fs pulses from the 270 fs output of a Light Conversion PHAROS laser (1024 nm, 400 \textmu J, 20 kHz). Spectral broadening occurs through self-phase modulation in the gas-filled HCFs, which is followed by temporal compression with chirped mirrors. The first stage (HCF1) yields \(\sim\)30 fs pulses and the second stage (HCF2) reduces it further to \(\sim\)15 fs. The latter pulse duration is in the regime required to drive soliton dynamics in a third HCF stage (HCF3).

\paragraph{HCF stages and pump generation}

The implementation of each HCF stage is equivalent and set up as follows. Each HCF stage consists of a mechanically stretched silica capillary fiber suspended in a gas cell. The input beam is focused into the waveguide through a telescope which enables tuning of the optimum focal spot size and focal plane position. The input and output windows of the gas cell are made of 1 mm fused silica, except for the output window of HCF3, which was chosen to be 1 mm MgF\(_{2}\) to improve UV transmission and reduce material dispersion. The outputs of HCF1 and HCF2 are collimated with a thin lens before being compressed with a set of chirped mirrors. The output of HCF3 is used as a probe pulse source for TA spectroscopy with the exact beam path illustrated in Fig. \ref{fig:Setup02_Referencing_layout}. Long-term beam pointing stability of each stage is facilitated through a one-point stabilization system at each stage, where a small fraction of the focusing beam is picked off and imaged onto a camera. A home-built software controls a piezo-controlled mirror mount to compensate for beam displacements. In combination with the high pointing stability of the laser source and the stable environmental conditions of the laboratory, the employed one-point stabilization system was sufficient to achieve an excellent long-term stability for each HCF output. All transmissive optics (windows, lenses, beam sampler, etc.) for the NIR are anti-reflection coated to cover the respective bandwidth. The transmission performance of each stage is determined through the coupling efficiency (CE). The CE is defined as the total transmission of the HCF divided by the calculated transmission of the lowest order HE\(_{11}\) mode at the central wavelength, assuming a straight waveguide according to the Marcatili and Schmeltzer model \cite{marcatili1964hollow}. A perfect Gaussian input can then be coupled into the lowest-order mode with a maximum CE of 98.1\%, which is achieved for a 1/e\(^2\) diameter of \(\sim\)64\% the HCF diameter \cite{nagy2008flexible}. However, in practice, some energy can be coupled to other modes, but this contribution to the transmitted energy cannot be distinguished when measuring the total transmission.

\begin{figure}[t!]
\centering\includegraphics[]{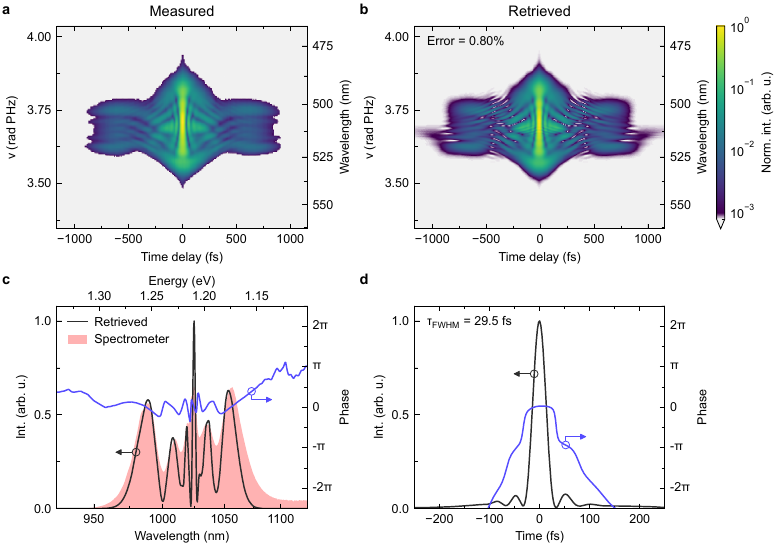}
\caption{SHG-FROG characterization of the output of post-compression stage HCF1. a,b) The measured and retrieved traces. c) The retrieved spectral intensity and phase. Also shown is the intensity corrected fundamental spectrum measured with a spectrometer. The spectral phase is shown upon propagation to the entrance of the next fiber. d) The retrieved temporal intensity and phase upon propagation to the entrance of the next fiber.}
\label{fig:FROG01_HCF1}
\end{figure}
\begin{figure}[t!]
\centering\includegraphics[]{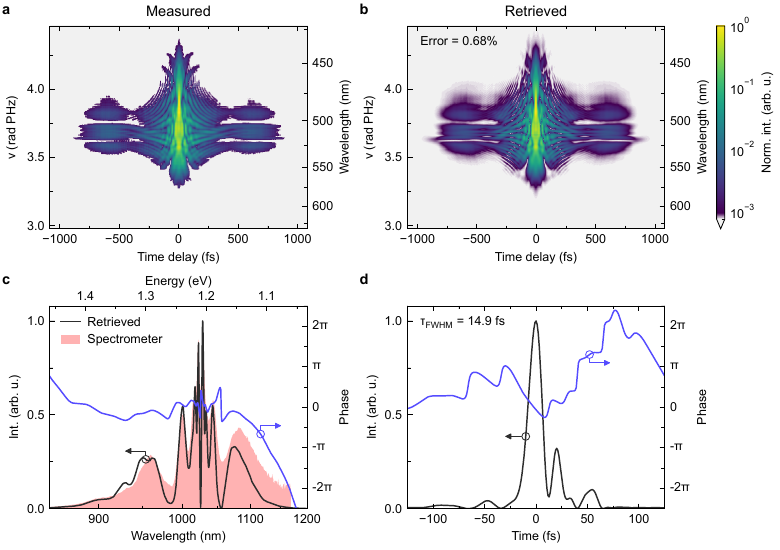}
\caption{SHG-FROG characterization of the output of post-compression stage HCF2. a,b) The measured and retrieved traces. c) The retrieved spectral intensity and phase. Also shown is the intensity corrected fundamental spectrum measured with a spectrometer. The spectral phase is shown upon propagation to the entrance of the next fiber. d) The retrieved temporal intensity and phase upon propagation to the entrance of the next fiber.}
\label{fig:FROG02_HCF2}
\end{figure}

Post-compression of the output of HCF1 and HCF2 is performed with several chirped mirrors of appropriate bandwidth (Ultrafast Innovations, HD120 and PC1611). The chirped mirrors following HCF2 impart a large negative chirp to compensate for the dispersion accumulated through the transmissive optics and several meters of air path between HCF2 and HCF3. Pulse compression is characterized through a home-built SHG-FROG. FROG traces from the measurements are retrieved using a home-built software implementation of the Principal Components Generalized Projections Algorithm (PCGPA). The obtained results are displayed in Supplementary Figs. \ref{fig:FROG01_HCF1} and \ref{fig:FROG02_HCF2}. The quality of the retrievals is excellent, as demonstrated by the very low FROG errors and the good correspondence between measured and retrieved traces and spectra. On this basis, we determine compressed pulse durations at full-width of half-maximum (FWHM) of 29.5 fs and 14.9 fs at the entrance of HCF2 and HCF3, respectively, in excellent agreement with the design targets (Supplementary Fig.~\ref{fig:Setup06_Target_design}).

A broadband beamsplitter (Layertec, 106102, R/T = 30\%/70\%) after HCF2 splits the beam into a pump and a probe beam, where the 70\% fraction is required to generate the supercontinuum probe in HCF3. The 30\% reflective fraction is used to generate the pump for TA measurements. The pump passes a 30 cm optical delay line (Aerotech, ATS115) and is focused through an optical chopper (Thorlabs) to reduce its repetition rate to 10 kHz. In addition to directly using the NIR pulse as a pump, excitation pulses at 510 nm and 340 nm can be obtained via second (SHG) and third harmonic generation (THG), respectively. For SHG, a 100 \textmu m thick type-I BBO is employed under loose focusing conditions and dispersion compensation is performed via a chirped mirror pair (Venteon Optics, DCM9). For THG, an additional 60 \textmu m thick type-II BBO is placed immediately after the SHG BBO and dispersion compensation is performed via a prism compressor. In all cases, the pump is focused onto the sample via a concave mirror and its linear polarization is controlled via a motorized HWP.

The 70\% fraction for the probe passes through a variable attenuator, consisting of a half-wave plate (Thorlabs, AHWP10M-980) and calcite Glan-Laser polarizer (Bernhard Halle Nachfl.), which allows adjustment of the pulse energy in HCF3 to tune the soliton fission length. The pressure in HCF3 is tuned for each experiment but is generally around 5.9 bar. Based on the small core diameter of 150 \textmu m and short length of 0.75 m, higher order modes other than the fundamental HE\(_{11}\) mode are expected to display negligible transmission efficiencies. In this respect, the calculated efficiencies for the three lowest order modes are T\(_{1024\text{ nm}}\)(HE\(_{11}\)) \(\approx\) 45\%, T\(_{1024\text{ nm}}\)(HE\(_{12}\)) \(\approx\) 1.4\% and T\(_{1024\text{ nm}}\)(HE\(_{13}\)) \(\approx\) 0.003\%. Indeed, the output profile on a card looks close to perfectly radially symmetric, reflecting the lowest order HE\(_{11}\) mode.

\paragraph{Probe pulse detection}

The probe path after HCF3 is illustrated in Fig. \ref{fig:Setup02_Referencing_layout}. A variable aperture is placed after the output window to spectrally attenuate the generated supercontinuum, as described in detail in Section \ref{sec:SI_DivergenceFocusing}. Afterwards, the transmitted beam is reflected off a wedged CaF\(_{2}\) window with a reflectivity of approximately 3\% to further attenuate the total probe pulse energy. The window also serves as a broadband beamsplitter for obtaining a reference beam from the reflection off its back surface. The shallow wedge angle spatially separates the two reflections. For TA measurements, the probe passes through the sample while the reference beam does not. The probe and reference spectra are detected simultaneously on a shot-to-shot basis via the two channels of a fiber-coupled, grating-based imaging spectrograph (Chromex 250IS). Detection is performed via a home-built CMOS-based double-chip detector. Due to the limited size of the employed chips (Hamamatsu, S10453-512Q), TA measurements are recorded over two separate spectral regions: 1) the DUV spanning 200--400 nm (DUV TA), and 2) the visible region spanning 400--800 nm (Vis TA). Each region is recorded via a dedicated grating (Richardson Gratings, DUV: 53009BK01-305H; visible: 53009KB01-471H). For measurements in the visible region, a 400 nm longpass filter (Thorlabs, FGL400S) is placed in the spectrograph to prevent detection of the second order grating diffraction, as well as a neutral density filter to avoid detector saturation. 

\subsubsection{Comparison of experimental performance of HCF3 to simulated soliton dynamics} \label{sec:SI_Comparison}

The output spectrum of the soliton stage HCF3 is recorded with a commercial fiber-coupled spectrometer (Avantes, AvaSpec-ULS2048XL-RS-EVO) equipped with an integrating sphere (Artifex, SP50). A quantitative comparison of the measured spectrum to the numerical simulation of the soliton dynamics requires the absolute intensity (radiometric) calibration of the employed spectrometer. However, as a suitable traceable calibration standard was not available, we pursued an approximate calibration via the procedure explained in section \ref{sec:SI_SpectrometerCalibration}. While the obtained calibration does not reach the quality required for a quantitative comparison with the simulations, it is sufficient to characterize the performance of HCF3. 

\begin{figure}[t!]
\centering\includegraphics[]{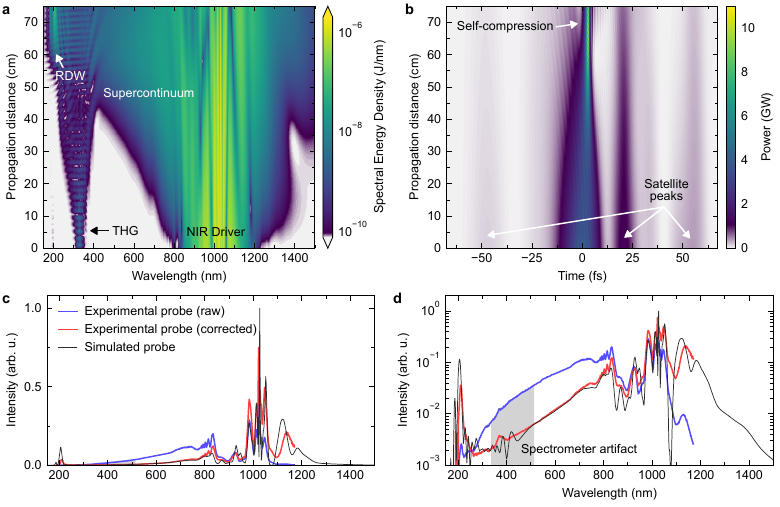}
\caption{Simulations of soliton dynamics in HCF3 using the experimentally retrieved pulse as input. a-b) The spectral and temporal evolution of the soliton upon propagation in the HCF. c-d) A comparison between the simulated spectral output and the experimental spectrum measured with a spectrometer. For the experiment, both the raw and corrected spectrum are shown. Spectra are scaled to visually match at the driver wavelength.}
\label{fig:Setup07_Simulations_HCF3}
\end{figure}

The soliton dynamics in HCF3 are simulated using the experimentally determined pulse parameters of the output of HCF2 as its input, and with helium at 5.9 bar as the filling gas. A good match with the experimental spectrum is achieved for an input pulse energy of 95 \textmu J, which is slightly lower than the experimental value that was estimated at 100 \textmu J. Fig. \ref{fig:Setup07_Simulations_HCF3}a,b shows the obtained spectral and temporal evolution upon propagation through the HCF. The output spectrum at the fiber exit is displayed in Fig. \ref{fig:Setup07_Simulations_HCF3}c,d along with the experimental spectrum. In the following, we first discuss the simulated soliton dynamics to identify the nonlinear processes contributing to its spectral characteristics. We then discuss how the simulations compare to the experiments.

\paragraph{Numerical simulation of soliton dynamics in HCF3}

Fig. \ref{fig:Setup07_Simulations_HCF3}a shows the spectral broadening of the soliton into a supercontinuum and the formation of the RDW at 206 nm. Fig. \ref{fig:Setup07_Simulations_HCF3}b shows that the main peak experiences self-compression at the same time as spectral broadening. The positions of satellite peaks from third order dispersion in the NIR input pulse are indicated in Fig. \ref{fig:Setup07_Simulations_HCF3}b. The simulation shows that upon propagation in the HCF, the satellite peaks are attenuated without experiencing self-compression. This implies that weak satellite structures in the NIR driver do not impact the soliton dynamics, which are exclusively driven by the main peak. We thus note that for the NIR driver pulse employed in HCF3, the measured total pulse energy overestimates the energy fraction driving the soliton dynamics, due to the energy fraction contained in its satellite structure.

As an additional contribution to the probe spectrum, the third harmonic of the NIR driver is generated throughout the HCF with its center near 330 nm as indicated in Fig. \ref{fig:Setup07_Simulations_HCF3}a. The simulation shows that the spectral bandwidth of the third harmonic increases along the fiber. This broadening can be attributed to two effects: 1) As the NIR driver broadens, so does the third harmonic that the driver creates, and 2) cross-phase modulation (XPM) with the intense NIR driver induces a broadening of the initially narrow third harmonic \cite{zheltikov1999self}. In the extreme case at the self-compression point, a very intense, octave-spanning and phase-locked third harmonic can be generated by the broadened soliton \cite{savitsky2023sub}. Consequently, at the fiber exit, the THG region spans approximately 160--500 nm, where it interferes with the soliton supercontinuum.

\paragraph{Comparison of simulated probe spectrum to experiment}

Figs. \ref{fig:Setup07_Simulations_HCF3}c-d show that the simulated soliton supercontinuum matches very well with the experimentally measured output of HCF3, once the spectral intensity calibration is taken into account. The strong deviation and increased noise above 1100 nm in the corrected spectrum is a consequence of the calibration procedure and explained in SI Section \ref{sec:SI_SpectrometerCalibration}. In the region of the RDW, the measured spectrum displays a substantially lower intensity below 210 nm compared to the simulations, and the RDW peak position appears slightly red-shifted. We suggest that this is in part caused by the employed calibration procedure, which is less accurate in this spectral region, similar to the NIR region. Additionally, small deviations in the exact experimental parameters, such as spectral phase, pulse energy and gas pressure, change the exact outcome of the simulations. Based on the good match with the experimental spectrum, one may estimate the spectral energy density for different regions within the calculated supercontinuum. Integration across the entire spectrum yields an energy of 43 \textmu J. The energy at the driver (850--1500 nm) is 35 \textmu J, for the supercontinuum between 220 and 850 nm it is 7.0 \textmu J, and at the RDW (160--220 nm) it is 1.1 \textmu J.

\subsubsection{Probe attenuation and focusing}\label{sec:SI_DivergenceFocusing}

For use in TA measurements, the overall probe pulse energy must be reduced to the nJ level, while the dominant contribution of the NIR region must be attenuated relative to the generated supercontinuum to achieve a uniform probe spectrum. However, due to the large spectral bandwidth, there are no suitable edgepass filters to retain the DUV while attenuating the visible to NIR regions. We therefore pursue an alternative spectral attenuation strategy that makes use of the fact that the beam divergence angle of the HCF output scales proportionally with the wavelength \cite{nagy2008flexible}. Consequently, a variable aperture placed after the HCF output can be used to predominantly cut the visible to NIR range, while leaving the DUV range mostly unaffected. Here we characterize the spectral attenuation as a function of aperture size and determine the focal spot sizes obtained with and without the aperture.

Beam profiles are measured with a beam profiler (Ophir Optronics, SP932U) and processed via its accompanying software BeamGage. Individual wavelength ranges are selected through bandpass filters (214 nm, Newport, 10BPF10-214; 270 nm, Newport, 10BPF10-270; 351 nm, Thorlabs, FBH351-10; 515 nm, Thorlabs, FBH515-10; 700 nm, Thorlabs, FBH700-10; 1030 nm, Thorlabs, FLH1030-10) and the associated beam diameters (1/e$^2$) along their major and minor axes are determined from 1D Gaussian fits.

\begin{figure}[t!]
\centering\includegraphics[]{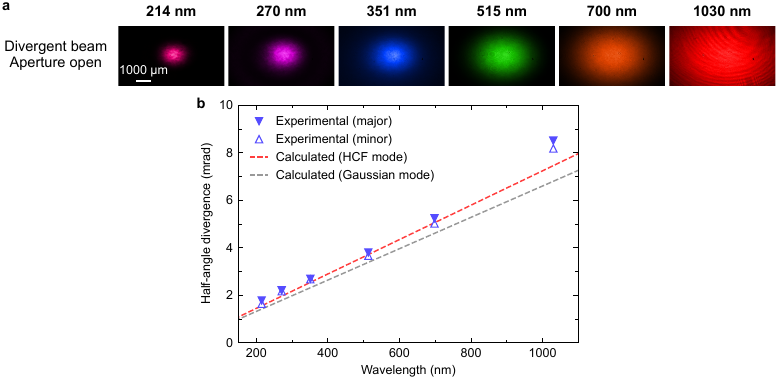}
\caption{Characterization of the output divergence of HCF3. a) Measured beam profiles for wavelength ranges selected with bandpass filters. b) Typical divergence of the major and minor axis for selected wavelengths in the continuum output. Also shown are the calculated divergence of the lowest order HE\(_{11}\) HCF mode from a fiber of diameter 150 \textmu m, and a Gaussian beam with a 1/e\(^{2}\) diameter of 64\% the HCF diameter.}
\label{fig:Setup08_DivergentBeam}
\end{figure}

\paragraph{Spectral attenuation of the supercontinuum probe}

We first determine the divergence out of HCF3 by measuring the beam diameter at a fixed distance of 65 cm for probe wavelengths selected by the bandpass filters. Typical beam profiles are displayed in Fig. \ref{fig:Setup08_DivergentBeam}a, whereas the half-angle divergence of the major and minor axis is displayed in Fig. \ref{fig:Setup08_DivergentBeam}b. The data shows that the divergence increases approximately linearly with increasing wavelength. To rationalize this behavior, one may observe that each spectral component of the continuum has the same beam waist at the HCF exit, such that their angular divergence becomes wavelength dependent. For a quantitative comparison, the output divergence of HCF3 was calculated following the procedure outlined in \cite{nagy2008flexible}. Briefly, the lowest order mode in the HCF is the HE\(_{11}\) mode, whose intensity profile follows the square of a Bessel function. In the far-field the half-angle divergence can then be found to scale with wavelength \(\lambda\) as \(\Theta_{\text{HCF}}=\lambda/(1.8413a)\), where \(a\) the HCF core radius. In comparison, the divergence of a Gaussian beam scales as \(\Theta_{\text{Gaussian}}=\lambda/(\pi w_{0}) \approx \lambda/(\pi \cdot 0.64a)=\lambda/(2.01a)\), where \(w_{0}\) is the 1/e\(^{2}\) radius, which is assumed to be 64\% of the HCF radius \cite{nagy2008flexible}. The Gaussian approximation thus leads to a linear scaling as well, but underestimates the divergence. Fig. \ref{fig:Setup08_DivergentBeam}b displays both the divergence from a Bessel beam and the divergence resulting from the Gaussian beam approximation. As expected, the calculation for the Bessel mode matches the experimental data very well, with a near quantitative match below 800 nm.

\begin{figure}[t!]
\centering\includegraphics[]{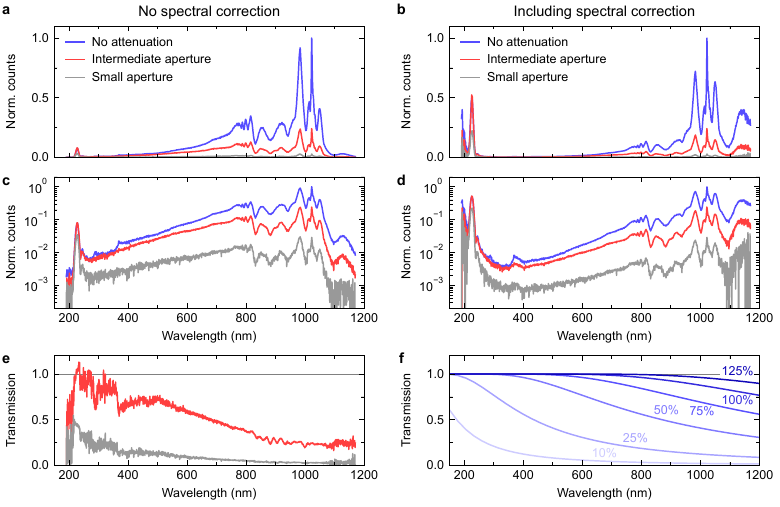}
\caption{Characterization of the probe's spectral attenuation with a variable aperture. a-d) HCF3 output without attenuation (blue solid line) and with attenuation via two different aperture sizes: an intermediate aperture size corresponding to \(\sim\)40\%, and a small aperture size corresponding to \(\sim\)10\% of the 1/e\(^{2}\) beam diameter at 1024 nm. The spectra are shown without (a,c) and with (b,d) an approximate spectral intensity calibration of the employed spectrometer. e) Transmission curves calculated from the measured spectra for both aperture sizes. f) Transmission curves calculated from the theoretical beam divergence of HCF3 for several aperture diameters given as the percentage of the 1/e\(^{2}\) diameter at 1024 nm.}
\label{fig:Setup05_Probe_attenuation}
\end{figure}

A variable aperture is placed in the divergent beam to radially crop the beam, thereby introducing a spectral attenuation that increases with the angular divergence. Fig. \ref{fig:Setup05_Probe_attenuation}a-d compares the unattenuated continuum spectrum with attenuated spectra from two different aperture diameters, displaying both the raw probe spectra (left panels) and the probe spectra corrected with the approximate spectral intensity calibration (right panels) described in Section \ref{sec:SI_SpectrometerCalibration}. The diameters of the intermediate and small aperture sizes correspond to \(\sim\)40\% and \(\sim\)10\% of the 1/e\(^{2}\) diameter at 1024 nm, respectively. Fig. \ref{fig:Setup05_Probe_attenuation}e displays the transmission curves calculated from the spectral data for each aperture size. The dips in transmission at 300 and 400 nm are attributed to a measurement artifact from the employed spectrometer caused by a small portion of the NIR signal leaking through to the detector in the 300--400 nm region. Consequently, closing the aperture reduces the artifact signal, leading to a dip in the transmission curves. Fig. \ref{fig:Setup05_Probe_attenuation}f shows theoretical transmission curves calculated for a Gaussian beam with the theoretical angular divergence obtained from a Bessel output mode from HCF3 (see Fig. \ref{fig:Setup08_DivergentBeam}b). The transmission is given for different aperture diameters relative to the 1/e\(^{2}\) diameter at 1024 nm.
 
Generally, the data shows that the aperture provides a straightforward method for improving the uniformity of the probe spectrum and reducing the intensity of the transmitted NIR driver to the level of the RDW emission. For the intermediate aperture size, this is achieved without perturbing the spectral intensity in the DUV below 400 nm. However, considering a detection window of 200--800 nm, the contrast between the lowest intensity at about 350 nm and the highest at 800 nm still corresponds to a factor of 10. This can be reduced to approximately a factor of 5 for the smallest aperture. However, this small aperture size also reduces spectral intensity in the DUV region. Furthermore, the experimental transmission curves are reproduced well by the theoretical calculations displayed in Fig. \ref{fig:Setup05_Probe_attenuation}f, highlighting that the impact of the spectral filtering aperture can be reliably estimated.

\begin{figure}[t!]
\centering\includegraphics[]{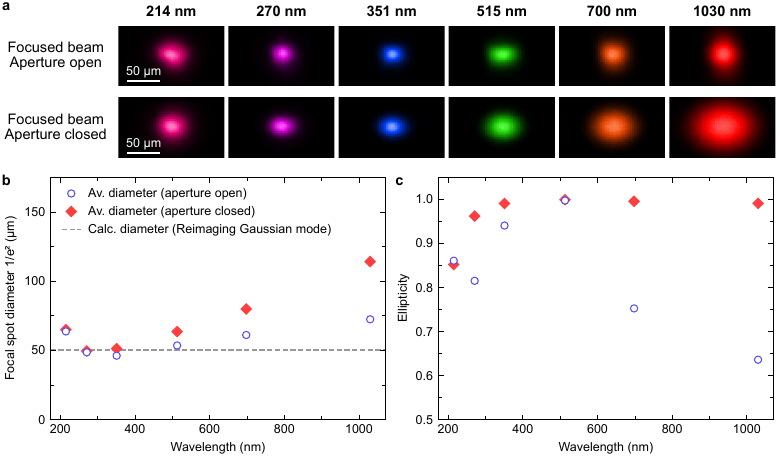}
\caption{Characterization of the focusing of the continuum output of HCF3. a) Measured beam profiles for wavelength ranges selected with bandpass filters. b,c) Focal spot diameter and ellipticity of the beam profiles shown in (a). The dotted line in (b) indicates the calculated 1/e\(^{2}\) diameter of a Gaussian beam that is re-imaged from the HCF output, considering the applied focusing optics and distances.}
\label{fig:Setup09_Focus}
\end{figure}

\paragraph{Characterization of probe focusing}

Following the aperture, the probe beam is refocused onto the sample position via a concave mirror. To determine the impact of the aperture on the focusing of the probe beam, Fig. \ref{fig:Setup09_Focus}a displays the focal spot profiles of several probe spectral regions selected via bandpass filters for two configurations: a fully open aperture that does not crop the probe beam (top row), and a closed aperture with an intermediate aperture size (\(\sim\)60\% of the 1/e\(^{2}\) diameter at 1024 nm, bottom row). Figs. \ref{fig:Setup09_Focus}b-c then display the average focal spot diameter and ellipticity extracted from Gaussian fits to the measured beam profiles. For the uncropped probe beam, a uniform average focal spot diameter in the range 50--70 \textmu m is obtained, which closely corresponds to the theoretical focal spot diameter calculated via a Gaussian beam re-imaged from the HCF exit (dashed line in Fig. \ref{fig:Setup09_Focus}c). Deviations from the theoretical value are most pronounced above 600 nm, where a substantial decrease in ellipticity is observed. This indicates the presence of astigmatism in the beam, which most likely results from the employed concave mirror. Closing the aperture then has two effects. First, the focal spot diameters of cropped spectral regions increase compared to the uncropped beam, which is mainly due to the associated reduction in effective angular divergence. Second, the ellipticity of cropped spectral regions is essentially maintained at unity, even for initially elliptical beam profiles. The aperture thus counteracts any astigmatism in the cropped spectral regions of the probe beam.

\subsubsection{Approximate spectral intensity calibration} \label{sec:SI_SpectrometerCalibration}

The equipment for measuring the ouput spectra of the HCF stages consist of a fiber-coupled spectrometer (Avantes, AvaSpec-ULS2048XL-RS-EVO), a multimode fiber (Thorlabs, FG200AEA or FG400AEA) and an integrating sphere (Artifex, SP50). Here we determine the spectral response of each of the three elements to achieve an approximate intensity calibration for the spectral measurements.

The spectral sensitivity of the Avantes spectrometer is estimated by measuring the spectrum of single wavelengths selected from the monochromator of a UV-Vis spectrometer (Shimadzu, UV-2600), and dividing the measured integrated intensity by the power recorded from a power meter (Thorlabs, S120VC). The power meter's sensitivity limits the accessible range to 200--1100 nm. To extend the calibration beyond that range, the sensitivity curve is exponentially extrapolated. The relative spectral response of the integrating sphere (Artifex, SP50) is determined from a measurement of a HCF3 continuum spectrum with and without the sphere. First the spectrum is measured with the sphere, which is connected to the spectrometer through an optical fiber. Second, the spectrum is measured by focusing directly into the optical fiber. The resulting ratio of the spectra then indicates the relative spectral efficiency of the integrating sphere. Generally, we find that the integrating sphere only relatively attenuates the spectrum below 400 nm. The inverse of the measured transmission efficiency can be approximated by an exponential function, which thus removes any noise from the measurement. Here, a continuum spectrum extending to 225 nm is used for the calibration, which is extended to shorter wavelengths through extrapolation of the exponential function. The last element in the detection is the spectral transmission of the optical fiber as well as their solarization-induced losses. Loss specifications were obtained from the supplier's website.

The range of spectral sensitivity dependence of the spectrometer is significantly larger than for the integrating sphere and optical fibers. The spectrometer's sensitivity is within one order of magnitude between 220 and 1060 nm but increases steeply outside that domain, such that measurement noise and minute baseline offsets can be amplified by the associated spectral correction. Therefore, spectral calibrations are only applied for the direct comparison between experimental and theoretical results and between measured and FROG retrieved spectra, whereas it is avoided otherwise. The NIR dependence is of particular importance when assessing retrieved FROG spectra, as the measured spectra get significantly biased to the high-energy portion in the absence of corrections.

\clearpage
\subsection{Characterization and suppression of measurement noise} \label{sec:SI_Noise}

\subsubsection{Noise Analysis of TA Measurements}

The measurement sensitivity of a TA setup is limited by multiple noise sources impacting the spectral detection, which have recently been reviewed by Lang \cite{lang2018photometrics}. Regarding the spectral detection, the dominant sources of noise include the intrinsic intensity fluctuations of the continuum probe, the detection process itself, including read-out and shot noise, and environmental fluctuations in the laboratory. Shot-to-shot detection combined with rapid data acquisition schemes, including optical chopping, rapid delay scanning, and high-repetition-rate laser sources, can improve data quality by mitigating low-frequency noise ($1/f$-noise) of the pump and probe pulses \cite{gueye2016broadband,draeger2017rapid}. Finally, referencing schemes can directly measure and correct the remaining intrinsic shot-to-shot intensity fluctuations of the probe continuum, such that the measurement sensitivities of state-of-the-art TA setups are only limited by the counting statistics of their detection system \cite{anderson2007noise,dobryakov2010femtosecond}. On this basis, the best reported setups currently reach a resolution of approximately 9~\textmu OD in 1 second of measurement time \cite{lang2018photometrics}. Here, we review the spectrally-resolved referencing schemes implemented in the HCF-based TA setup, following the noise analysis by Lang\cite{lang2018photometrics} and the Ge group \cite{feng2017general,feng2019optimized}.

In the absence of referencing, the standard deviation of a TA measurement is provided by

\begin{equation}
\label{eqn:Noise_nonRefODNoise}
\sigma_{\Delta A} = \frac{1}{\ln{(10)}} \sqrt{\frac{2}{S^2}\sigma_S^2 + \kappa^2 \sigma_{I_\text{ex}}^2}.
\end{equation}

\noindent where $S$ is the measured probe signal including detector noise:

\begin{equation}
\label{eqn:Noise_nonRefProbe}
\frac{2}{S^2}\sigma_\text{S}^2 = \frac{2}{I_\text{S}^2}\sigma_{I_\text{S}}^2 + \frac{2}{S^2}\sigma_{\text{det}}^2.
\end{equation}

\noindent Here $I_S$ is the probe intensity without detection and $\sigma_{\text{det}}$ is the noise arising from the detection (read-out, shot and digitization noise). These equations assume weak pump-induced signals such that the pumped ($S^*$) and unpumped ($S^0$) probe signals are approximately equal ($S^* \approx S^0$) and both are denoted as $S$. $I_\text{ex}$ is the pump intensity and $\kappa I_\text{ex}$ defines the pump-induced transient absorption. The first term in \autoref{eqn:Noise_nonRefODNoise} thus describes probe-related noise, including both white light intensity fluctuations and detector noise, while the second term accounts for pump-induced noise. Optimal performance is achieved by minimizing probe, pump, and detector noise, while maximizing the detected signal $S$. In the absence of a pump pulse or if pump intensity noise is negligible, one obtains that the standard deviation of a TA measurement scales linearly with the standard deviation of the detected probe signal intensity including detector noise: 

\begin{equation} \label{eqn:Noise_nonRefProbeOnly}
    \sigma_{\Delta A} \text{[mOD]} =6.1\sigma_\text{S} \text{[\%]}.
\end{equation}

\noindent Therefore, 1\% noise in the detected probe signal corresponds to a TA standard deviation of 6.1~mOD. For shot-resolved detection, averaging must be performed carefully to avoid introducing artifacts, which can distort the TA spectrum from its true value \cite{brazard2015accurate}. In this respect, the most robust scheme for averaging non-referenced single-shot data is:

\begin{align}
\label{eqn:Noise_nonRefAve}
\Delta A = - \log_{10} \left(\frac{\langle S^*\rangle}{\langle S_{0}\rangle}\right),
\end{align}

\noindent where $\langle...\rangle$ denotes the average.

\paragraph{B-matrix referencing}

Recently, the Ge group developed a generalized noise suppression scheme that takes advantage of the full spectral correlations between the probe and reference continua\cite{feng2017general,feng2019optimized}. In this correlation matrix or \textit{B-matrix} approach, intensity noise of the supercontinuum is corrected using a subtractive scheme based on the difference between consecutive shots. Here, the calculation is performed with arrays of signal ($\mathbf{S}$) and reference ($\mathbf{R}$) intensities, where each element in the array corresponds to the measured intensity at a single pixel:

\begin{align}
\label{eqn:Noise_bMatrixDiff}
\Delta \mathbf{S} = \mathbf{S}^*-\mathbf{S}^0\ \ \ \Delta \mathbf{R} = \mathbf{R}^*-\mathbf{R}^0.
\end{align}

\noindent The corrected pumped-unpumped difference in this subtractive scheme is provided by 
\begin{align}
\label{eqn:Noise_bMatrixK}
\Delta\mathbf{K} &= \Delta\mathbf{S} - \mathbf{B}\Delta\mathbf{R}.
\end{align}

\noindent Here, the B-matrix defines the optimal linear combination of reference pixels used to correct each probe pixel:

\begin{align}
\label{eqn:Noise_bMatrix}
\mathbf{B} = \text{cov}(\Delta\mathbf{R})^{-1}\text{cov}(\Delta\mathbf{R},\Delta\mathbf{S}),
\end{align}

\noindent where $\text{cov}()$ is the covariance matrix. This approach minimizes the variance of the corrected signal based on the spectral correlations of a series of probe and reference shots across all wavelengths. This differs from ratiometric pixel-to-pixel referencing\cite{dobryakov2010femtosecond}, which only relies on one-to-one pixel matching of a pair of individual probe and reference shots assuming the maximum positive correlation. In contrast, the B-matrix is determined from a separate measurement without pump excitation and weights all reference pixels to correct noise in each signal pixel. In this way, the B-matrix provides the optimal match to suppress the measured probe intensity fluctuations via the reference and is not limited by the probe-reference correlations between individual probe and reference pixel pairs.

The transient absorption signal in the B-matrix method is obtained using

\begin{equation}
\Delta \mathbf{A} = \frac{2}{\ln(10)}\frac{\Delta \mathbf{S}-\mathbf{B} \Delta \mathbf{R}}{\mathbf{S}^* +\mathbf{S}^0},
\label{eqn:Noise_bMatrixOD}
\end{equation}

\noindent where a linear approximation of the logarithm is applied. Since the equation is linear, the order of averaging has no effect such that a robust convergence to an artifact-free TA spectrum is ensured. As the B-matrix method exploits spectral correlations that are not utilized in the traditional one-to-one ratiometric referencing scheme\cite{dobryakov2010femtosecond}, it provides superior noise suppression.

In this correlation matrix method, the detector noise floor is not straightforward to determine because it depends on the quality of the B-matrix, which is estimated from a separate measurement. In the limiting case where all elements of the B-matrix are zero (no correlation between probe and reference signals), the noise floor of the non-referenced detection scheme is recovered since only one detector contributes to the TA signal. Similarly, in the limiting case where the determination of the B-matrix converges to the theoretical optimum, B-matrix referencing approaches the noise floor of the probe signal detector \cite{feng2019optimized}.

\subsubsection{Noise statistics of NIR driver and SHG pump}

\begin{figure}[!t]
\centering\includegraphics[width=\textwidth]{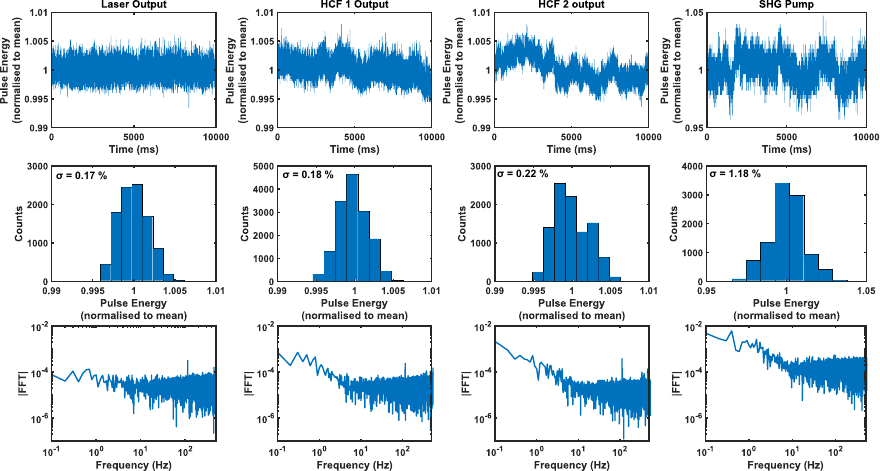}
\caption{Noise Statistics for the laser output, HCF1 output, HCF2 output and SHG pump (left to right). Top row: Pulse energy as a function of time. Middle row: Histograms of pulse energy with the percentage standard deviation indicated in the top left. Bottom row: FFT of the pulse energy.}
\label{fig:Noise_SI_01_stats_fund}
\end{figure}

Figure~\ref{fig:Noise_SI_01_stats_fund} displays the stability and noise characteristics of the Pharos output, HCF1, HCF2, and the SHG pump over 10~seconds at a 1~kHz sampling rate, limited by the energy meter (Coherent, J-10MB-LE). The laser exhibits low shot-to-shot noise (0.17\%) with a flat frequency spectrum aside from a small component at 115~Hz. After propagation through HCF1 and HCF2, the relative intensity noise increases slightly (0.18\% and 0.22\%, respectively) and a clear $1/f$-noise component emerges with a $1/f$-corner around 10~Hz, along with the persistent 115~Hz feature. Since self-phase modulation in HCFs redistributes spectral energy and consequently only broadens the spectrum, the total (spectrally-integrated) energy noise is mostly preserved with only minor increases. Despite no significant alteration in the energy noise, the emergence of low-frequency components suggests slow fluctuations in pulse energy after propagation through the fibers. This $1/f$-noise likely arises from pointing fluctuations, which marginally increase energy noise. 

The SHG pump shows a substantially higher noise level (1.18\%), representing an amplification of about 5.4~times compared to the output of HCF2. Although SHG is expected to double intensity noise due to its quadratic dependence on input intensity, the observed increase exceeds this expectation, most likely due to increased detector noise at lower pulse energies ($\approx$~3 \textmu J). We therefore expect that the measured noise level of 1.18\% represents an upper limit imposed by the detection method. The SHG output retains the $1/f$-noise behavior with a similar 10~Hz corner frequency. However, we here note that in TA measurements, pump fluctuations are typically negligible compared to fluctuations in the probe. In the one-photon excitation regime, TA noise arising from pump fluctuations scales approximately with the TA signal (Equation \ref{eqn:Noise_nonRefODNoise}). For a pump intensity noise of 1~\%, a TA signal of approximately 100~mOD would be required for pump-induced noise to contribute at the same level as the probe noise achieved with the B-matrix scheme ($\approx$1~mOD). As the typical signal amplitudes in this work do not exceed 30~mOD, the contribution from pump noise can be considered negligible.

\subsubsection{Detector Noise}
\label{sec:SI_Noise_detector}

The Hamamatsu S10453--512Q sensor (15-bits) is employed for both probe and reference detectors. Considering the detector specifications \cite{detector_specs}, the TA noise floor for 100\% detector saturation is calculated to be 0.8~mOD for non-referenced and B-matrix referenced data. This is in agreement with Lang \cite{lang2018photometrics} where a detailed calculation procedure can be found. 

It is important to note that this analysis assumes that the detector noise is uncorrelated between the probe and reference detectors. However, we find a positive correlation between probe and reference detector signals in the absence of illumination. This is likely due to both detectors being controlled via a single electronic board and therefore exhibiting correlated read-out noise. Indeed, it can be observed in \autoref{fig:Noise_TA_Main}b that at wavelengths where detector counts are high (e.g. above 600~nm), the experimental standard deviation is below the calculated noise floor of 0.8~mOD.  We therefore suggest that the detector noise correlation between probe and reference allows partial correction of the read-out noise.

\clearpage
\subsubsection{Probe Intensity Noise}

\begin{figure}[htbp]
\centering\includegraphics[width=\textwidth]{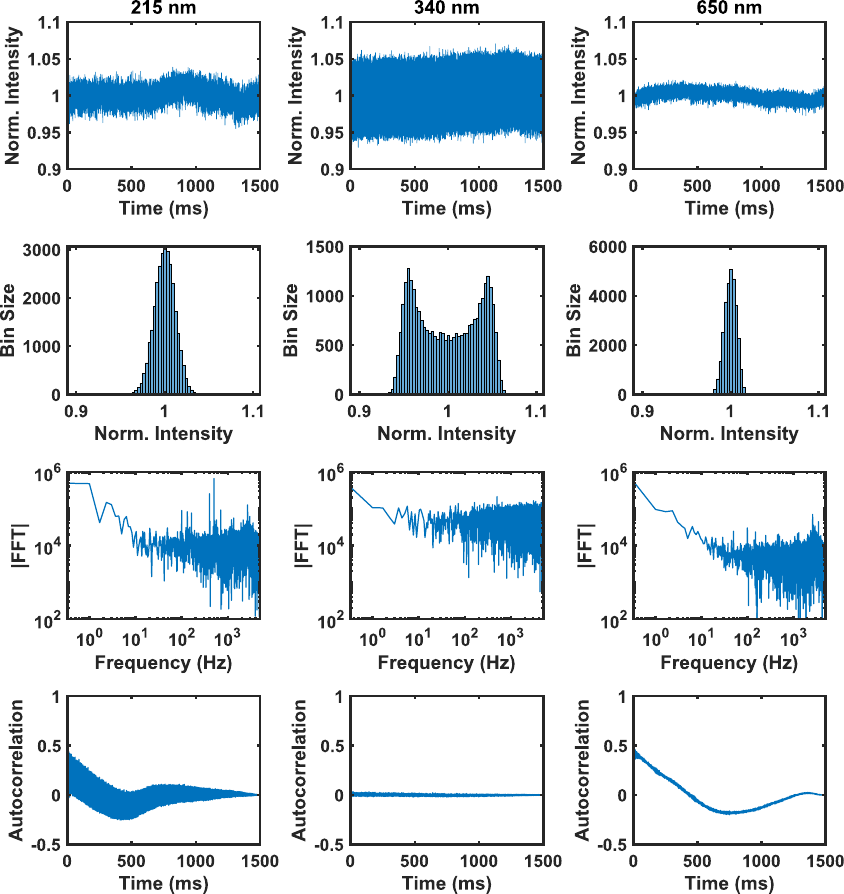}
\caption{Probe intensity statistics at 215~nm (left), 340~nm (middle) and 650~nm (right). First row; trajectory, second row; histogram, third row; FFT, fourth row; autocorrelation. The trajectories and histogram intensity values are normalized to the mean intensity.}
\label{fig:Noise_SI_Intensity_stats}
\end{figure}

\begin{figure}[htbp]
\centering\includegraphics[width=\textwidth]{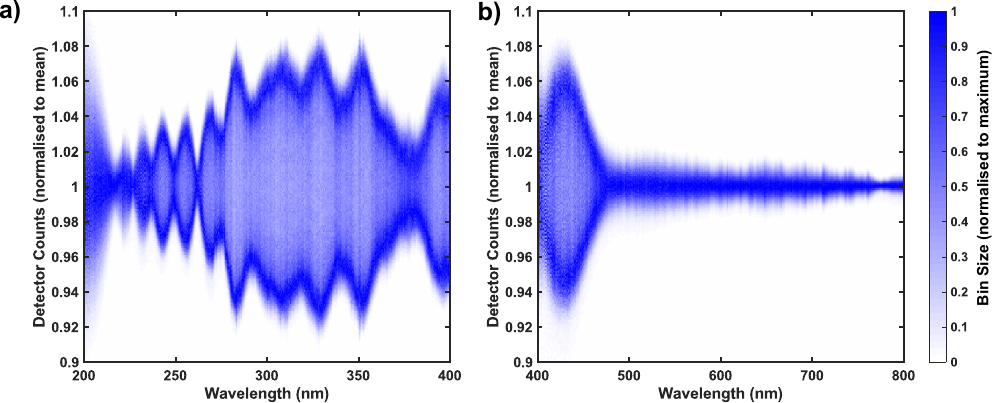}
\caption{Contour plot of the histogram of the normalised intensity at every wavelength in the \textbf{a)} DUV and \textbf{b)} visible regions. The bin size at each wavelength have been normalized to its maximum value to facilitate the comparison of different wavelengths.}
\label{fig:Noise_SI_all_histograms}
\end{figure}

\begin{figure}[htbp]
\centering\includegraphics[width=0.9\textwidth]{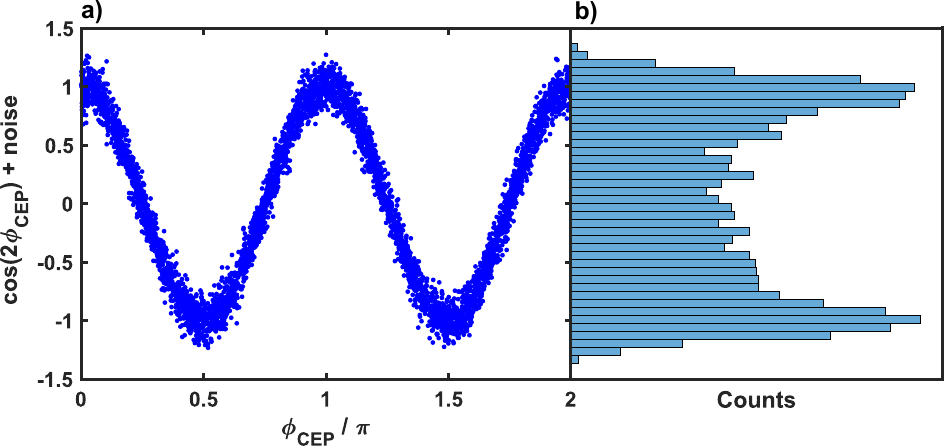}
\caption{Bimodal intensity distribution around the mean value arising due to the CEP-dependence of the interference between the soliton and THG. \textbf{a)} Cosine curve with added Gaussian noise. \textbf{b)} Histogram of the $\cos{(2\phi_\text{CEP})}$ with added Gaussian noise in \textbf{a} for randomly distributed $\phi_\text{CEP}$-values.}
\label{fig:Noise_SI_cos_Bimodal}
\end{figure}

\clearpage
\subsubsection{Spectral Correlations}

\begin{figure}[htbp]
\centering\includegraphics[width=\textwidth]{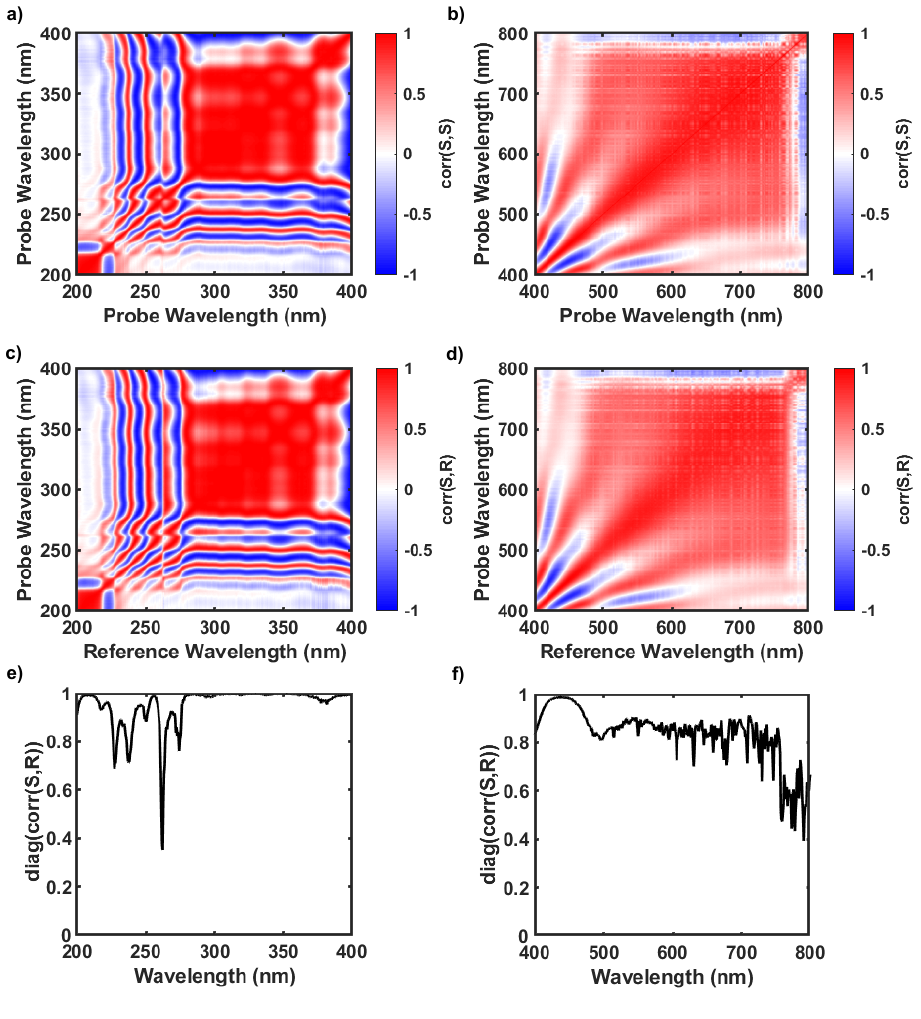}
\caption{Correlation matrices showing the Pearson correlation coefficient obtained from a series of probe (S) and reference shots (R). a) The probe-probe correlation matrix from 200 to 400~nm. b) The probe-probe correlation matrix from 400 to 800~nm. c) The probe-reference correlation matrix from 200 to 400~nm. d) The probe-reference correlation matrix from 400 to 800~nm. e) Diagonal elements of corr(S,R) from 200 to 400 nm. f) Diagonal elements of corr(S,R) from 400 to 800 nm.}
\label{fig:Noise_SI_02_corr_probe}
\end{figure}

\begin{figure}[htbp]
\centering\includegraphics[width=\textwidth]{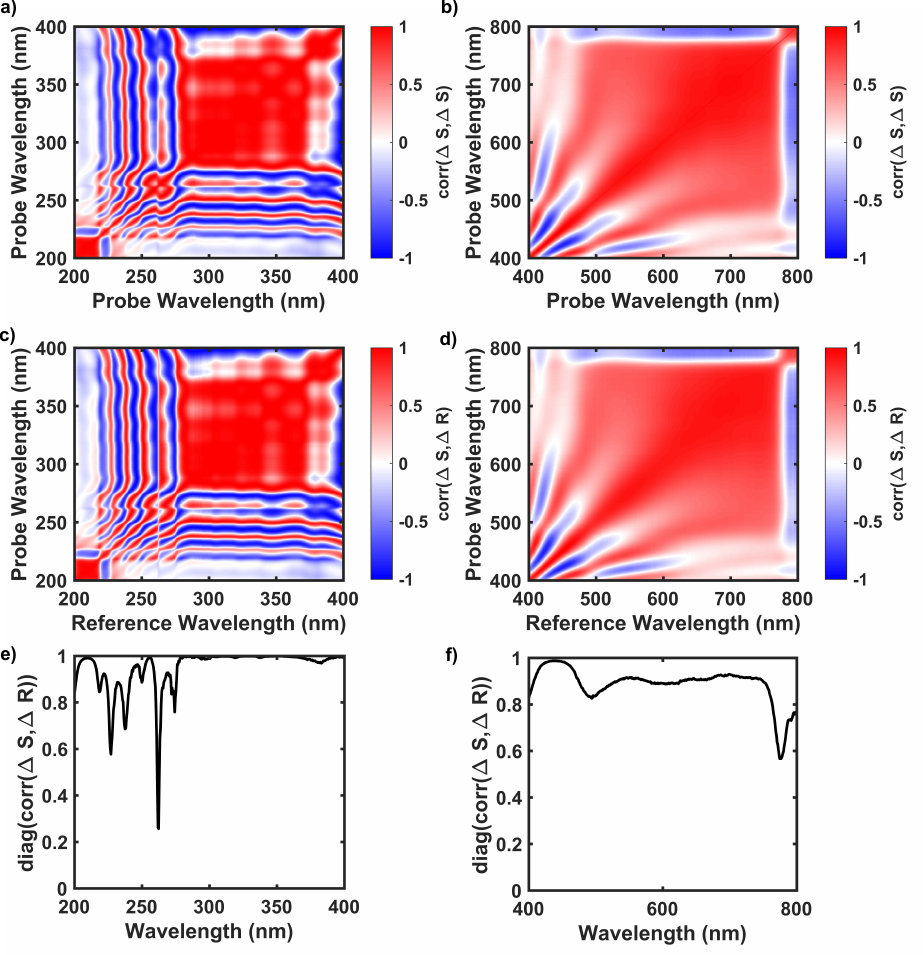}
\caption{Correlation matrices of the pairwise intensity differences of probe and reference. a) The probe-probe correlation matrix from 200 to 400~nm. b) The probe-probe correlation matrix from 400 to 800~nm. c) The probe-reference correlation matrix from 200 to 400~nm. d) The probe-reference correlation matrix from 400 to 800~nm. e) Diagonal elements of corr(\(\Delta\)S,\(\Delta\)R) from 200 to 400 nm. f) Diagonal elements of corr(\(\Delta\)S,\(\Delta\)R) from 400 to 800 nm. The correlation matrices of the pairwise intensity differences display the same correlation structures as the corresponding matrices in Fig.~\ref{fig:Noise_SI_02_corr_probe}. This suggests that the associated intensity noise is dominated by shot-to-shot fluctuations rather than low-frequency noise.}
\label{fig:Noise_SI_03_corr_shots}
\end{figure}

\clearpage
\subsubsection{Statistical Analysis of TA Signal}

\begin{figure}[htbp]
\centering\includegraphics[width=\textwidth]{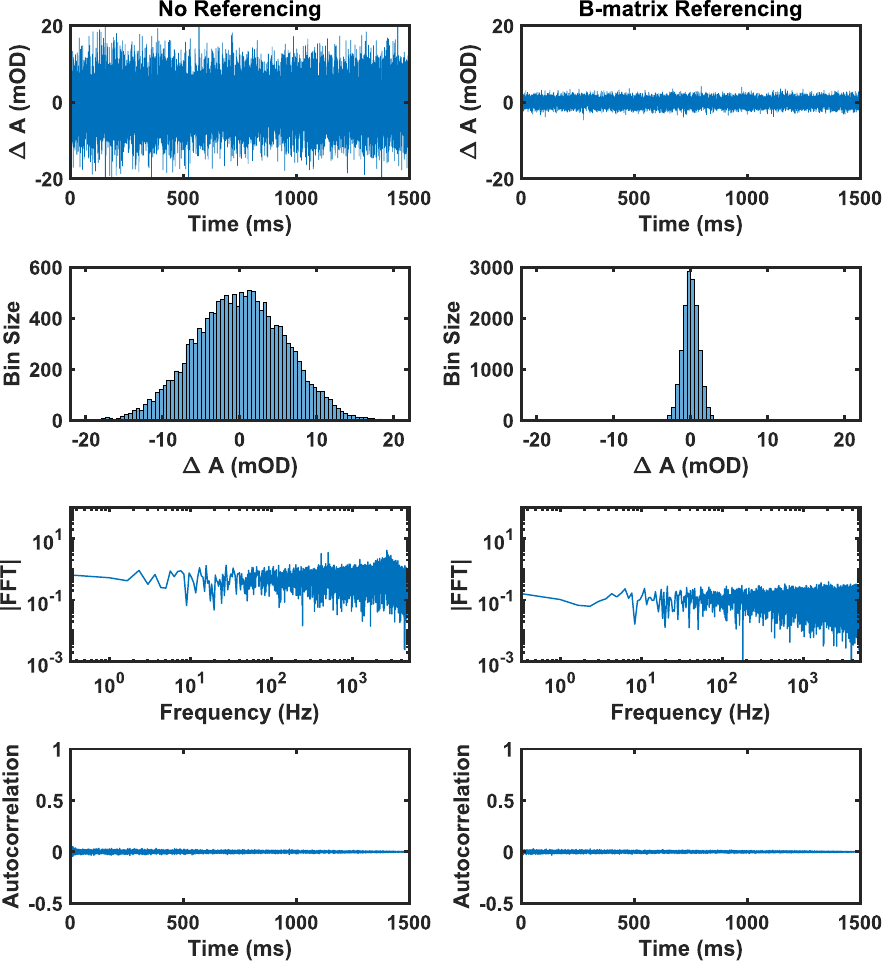}
\caption{TA statistics at 215~nm for no referencing and B-matrix referencing. First row; trajectory, second row; histogram, third row; FFT, fourth row; autocorrelation.}
\label{fig:Noise_SI_04_stats_215}
\end{figure}

\begin{figure}[htbp]
\centering\includegraphics[width=\textwidth]{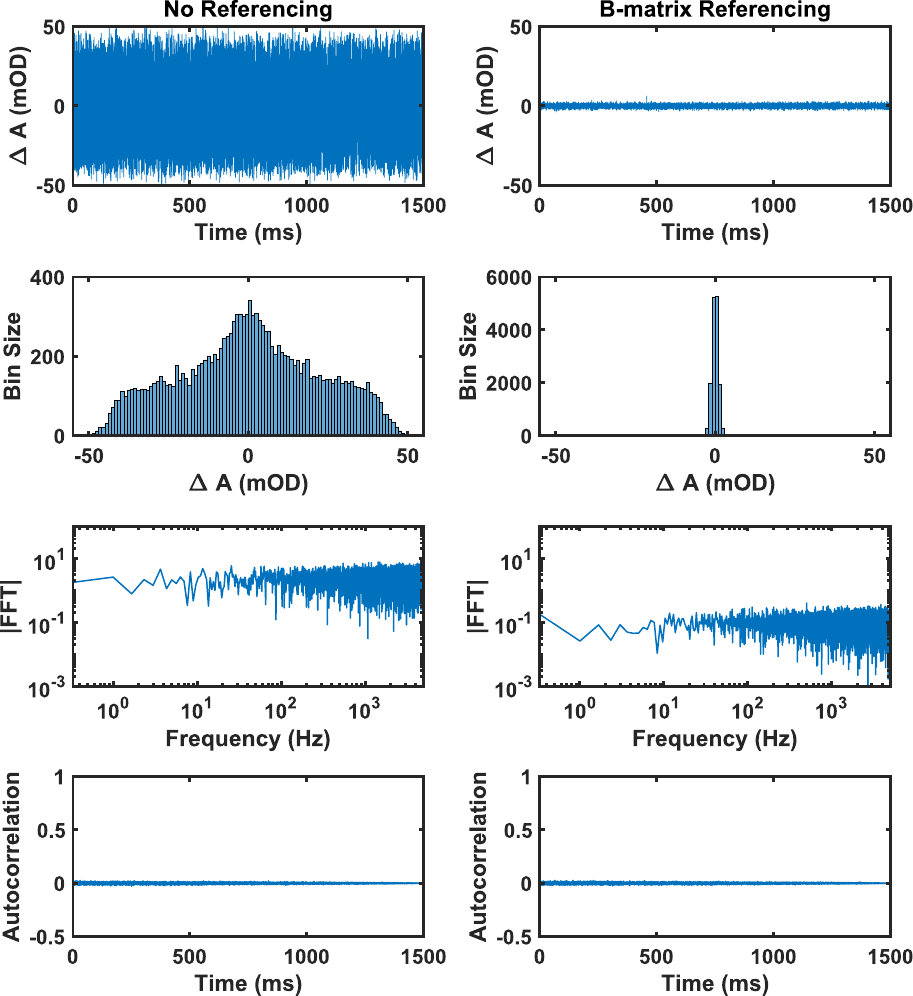}
\caption{TA statistics at 340~nm for no referencing and B-matrix referencing. First row; trajectory, second row; histogram, third row; FFT, fourth row; autocorrelation.}
\label{fig:Noise_SI_04_stats_340}
\end{figure}

\begin{figure}[htbp]
\centering\includegraphics[width=\textwidth]{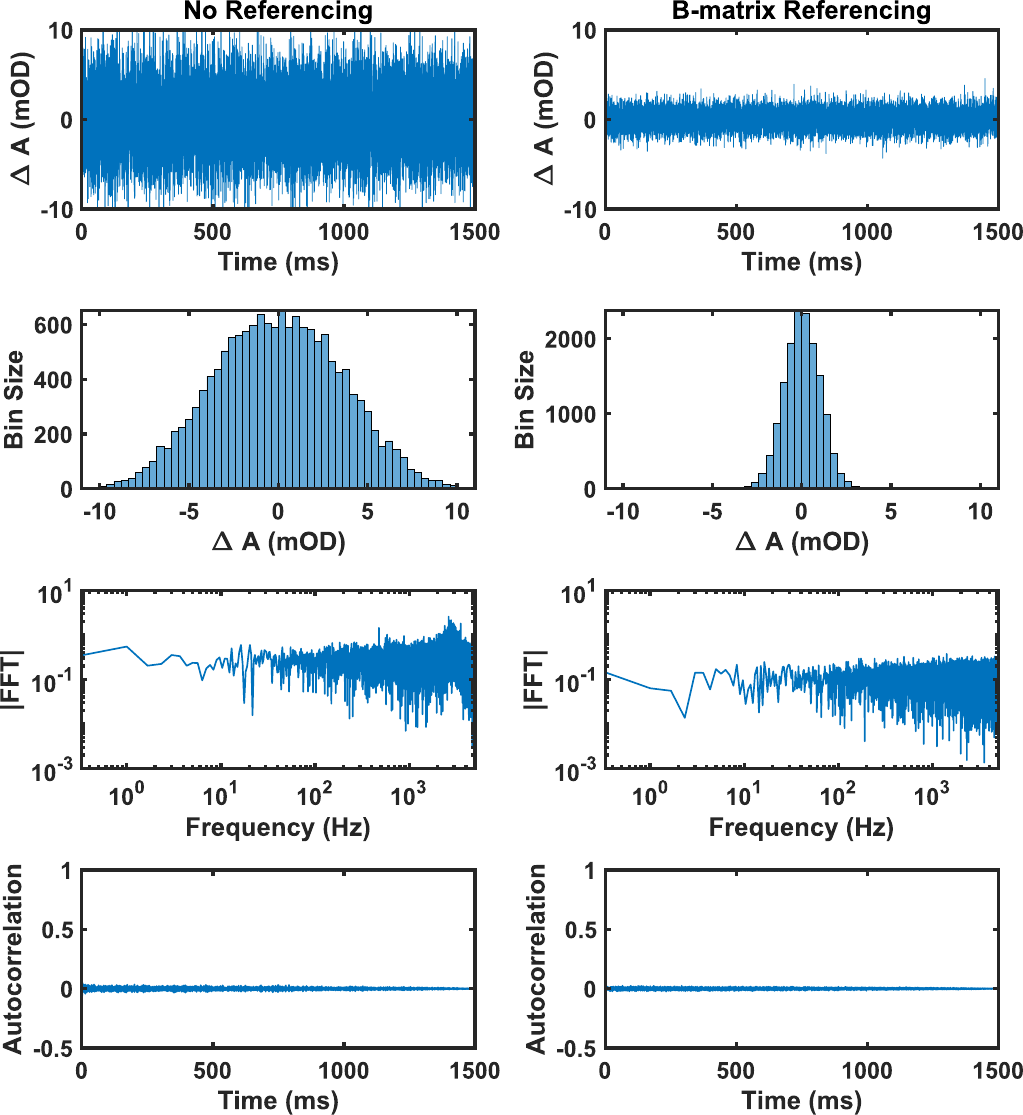}
\caption{TA statistics at 650~nm for no referencing and B-matrix referencing. First row; trajectory, second row; histogram, third row; FFT, fourth row; autocorrelation.}
\label{fig:Noise_SI_04_stats_650}
\end{figure}

\begin{figure}[htbp]
\centering\includegraphics[width=\textwidth]{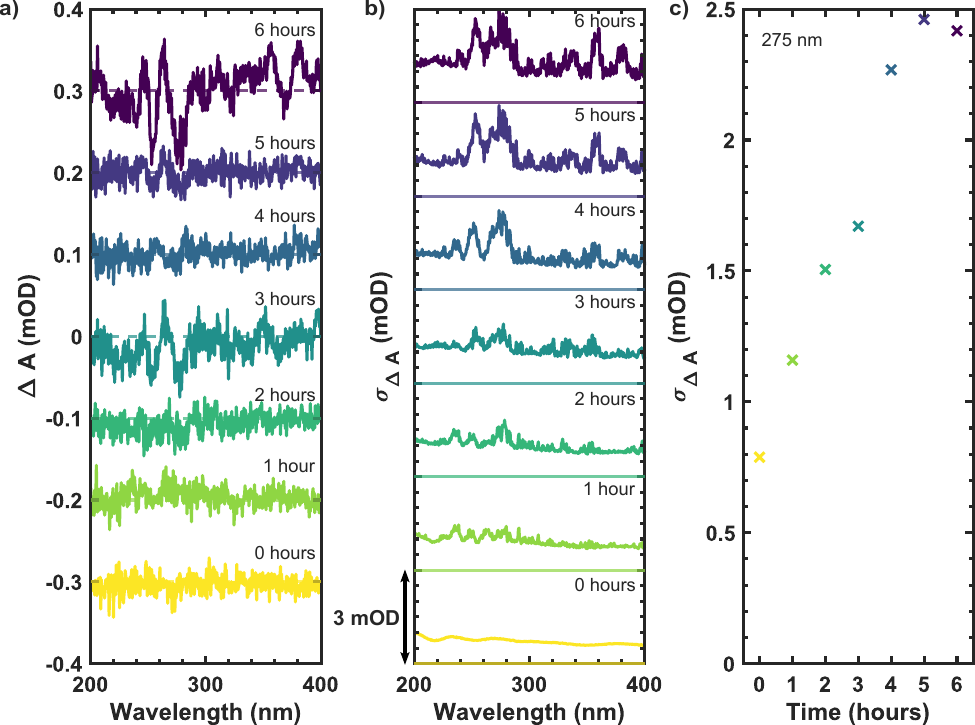}
\caption{Evaluation of the long-term validity of a B-matrix for referencing later TA measurements. A B-matrix is calculated at 0~hours and this same B-matrix is used to correct data each subsequent hour as indicated in the figure. a) Baseline measurement. Each measurement is offset vertically to aid visualization, with the dashed line of the same color indicating $\Delta \text{A} = 0$. b) Standard deviation. Each measurement is offset vertically to aid visualization with the solid line of each color indicating $\sigma_{\Delta \text{A}} = 0$. c) Standard deviation at 275~nm as a function of time.}
\label{fig:Noise_SI_05_bmat}
\end{figure}

\clearpage
\subsection{Ultra-broadband transient absorption}\label{sec:SI_Ultrabroadband_Transient_absorption}

\subsubsection{Time resolution and chirp analysis}\label{sec:SI_IRF}

The time resolution in TA measurements depends on various factors. For any broadband pump and probe, compressed or not, the effective response function (ERF) determines the fastest rate of change of any observable TA feature \cite{polli2010effective}. For example, the ERF limits the fastest signal rise or decay and the highest frequency of any coherent oscillation that may be present in the TA data. For the common case of a compressed pump and a broadband probe that only experiences linear chirp, the ERF is a convolution between the pump pulse duration and the pulse duration of the Fourier transform limit (FTL) of the probe \cite{polli2010effective,liebel2015principles}. In this respect, higher-order dispersion in the probe will temporally broaden the ERF, even though a simple functional relationship is not available \cite{polli2010effective}.

Independent of the ERF, the presence of any pulse induced artifacts around time zero, caused by the temporal overlap of the pump and probe pulses in the sample system, may limit the earliest pump-probe delays where a pure TA signal from the solvated sample is accessible \cite{beckwith2020data}. Here, the most important contributions arising from the interaction between the pump and probe are cross-phase modulation (XPM) in the flow cell windows and the solvent, and two-photon absorption in the solvent (2PA) \cite{beckwith2020data,ekvall2000cross,baudisch2018time}. Generally, one observes a superposition of the effects, which is typically referred to as a Coherent Artifact (CA) signal. While the relative strengths of the XPM and 2PA contributions depend on both the employed optical media and the pulse parameters, XPM will usually be present across the entire probe bandwidth, whereas 2PA contributions tend to be largest in the UV range where many solvents display increased 2PA cross sections. For a compressed pump and a linearly chirped probe, the temporal width of an XPM and a 2PA signal is generally determined by the wavelength-dependent cross-correlation between the pump and the continuum probe, thus providing a measure of the ERF \cite{kovalenko1999femtosecond}.

However, while isolated XPM or 2PA signals can provide an estimate for the effective time resolution of a TA measurement under favorable conditions, their contributions often overlap and are difficult to disentangle. Moreover, multiple XPM interactions within the same probe pulse can be observed from the front and back face of a flow cell due to temporal walk-off with respect to the pump along the interaction path \cite{ekvall2000cross}. Naturally, the temporal separation of the XPM contributions scales with the pump-probe walk-off, such that large differences in the group velocities of pump and probe and long interaction paths are expected to broaden the overall temporal width of the resulting CA. In this section, we provide a comprehensive analysis of these effects to determine the effective time resolution of the TA setup. In this respect, we place a special emphasis on the DUV region, where the nonlinear increase in material dispersion is expected to induce higher-order dispersion in the probe and increase any pump-probe walk-off present in the CA.

\paragraph{Measurement of coherent artifact in pure H\(_{2}\)O}

\begin{figure}[t!]
\centering\includegraphics{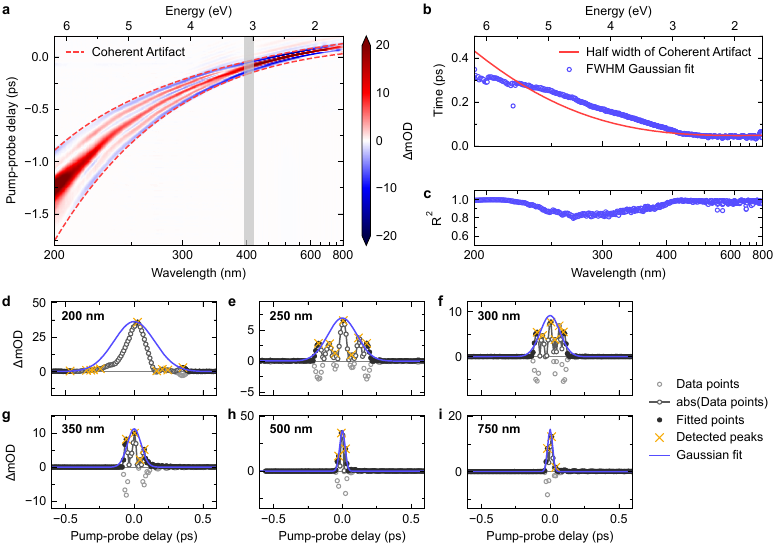}
\caption{The temporal width of the coherent artifact. a) The TA solvent response of H\(_{2}\)O. The dashed lines indicates where the signal of the CA is going to zero. b) Two measures of the width of the CA: half of the width between the dashed lines indicated in (a) and the FWHM of a Gaussian fit. c) The corresponding R\(^{2}\) value of the Gaussian fit. d-i) Exemplary traces to show the quality of the fit which yield the FWHM in (b).}
\label{fig:IRF01_Experimental_time_resolution}
\end{figure}

To determine the CA under typical measurement conditions, we here measure the TA signal of H\(_{2}\)O in a 100 \textmu m pathlength flow cell with 200 \textmu m thick thin-wall-aperture windows (Starna, 48/UTWA2/Q/0.1). The pump focal spot 1/e\(^{2}\) diameter is \(\sim\)240 \textmu m and the one of the probe varies between 50-80 \textmu m depending on the wavelength. The pump polarization direction is set at the magic angle (\(\sim 54.7^{\circ}\)) with respect to the probe. The obtained data is shown in Fig. \ref{fig:IRF01_Experimental_time_resolution}a. In the following, we first analyse the temporal width and chirp of the CA. Afterwards, the ERF is estimated from the data where 2PA signals can be isolated.

\paragraph{Coherent artifact and chirp analysis}

\begin{figure}[t!]
\centering\includegraphics{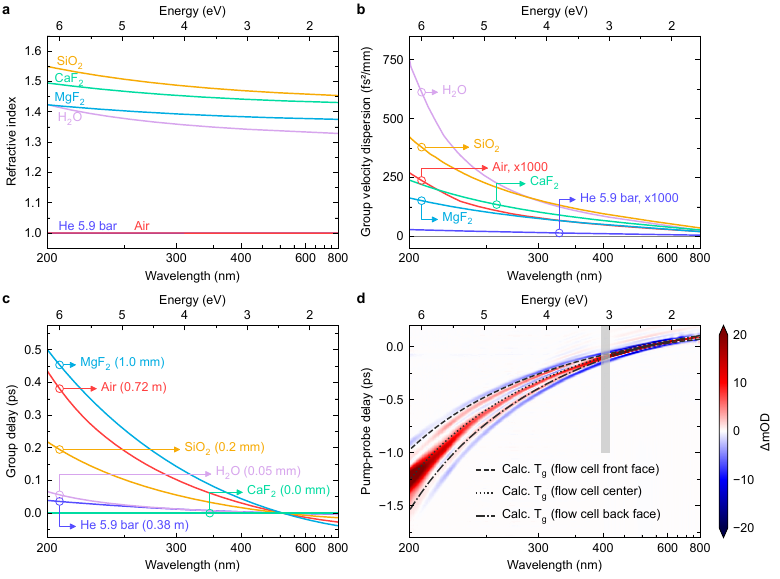}
\caption{Analysis of the experimental probe dispersion between the exit of HCF3 and the sample volume. a) The refractive index of materials employed in the path. b) The GVD calculated from the refractive index. c) The calculated group delay according to the pathlengths and gas pressure in the experimental setup up to the center of the flow cell. d) TA solvent response of H\(_{2}\)O. Overlayed are the inverted, calculated group delay at the front face, center and back face of the flow cell. The group delays in (c) and (d) are shown relative to the value at 510 nm.}
\label{fig:IRF02_Probe_chirp_analysis}
\end{figure}

Fig. \ref{fig:IRF01_Experimental_time_resolution}a displays the TA response of H\(_{2}\)O. The signal above 375 nm comprises only a few signal modulations, whereas in the UV several modulations occur with a strong positive signal appearing in the center below 250 nm. The near-Gaussian-shape of the positive signal below 250 nm is identified as 2PA, while the strong signal modulations are assigned to XPM contributions. Generally, the chirp of the CA increases nonlinearly in the DUV, such that the group delay between the probe's spectral components at 200 nm and 800 nm reaches approximately 1.2 ps at the center of the CA. Notably, the temporal width of CA displays a similar behavior and scales approximately with the observed chirp.

The width of the CA is quantified in Fig. \ref{fig:IRF01_Experimental_time_resolution}. The two dashed lines denote the full width of the CA, based on its signal going to zero. The half width of the masked interval is plotted in Fig. \ref{fig:IRF01_Experimental_time_resolution}b and provides an estimate of how much of a sample's early time dynamics will be obscured by the CA. Alternatively, the CA may be decomposed into a linear combination of a Gaussian function and its derivatives, thereby linking the ERF to the full-width at half-maximum (FWHM) of the fitted Gaussian \cite{kovalenko1999femtosecond,slavov2015implementation}. However, this approach is not suitable below 380 nm as the signal displays too many peaks and troughs. Therefore, an adapted version of a Gaussian fit is applied following the procedure reported in ref.\cite{oppermann2022chiral}. Here, the absolute value of the signal is taken and the data between its outermost peaks excluded, with the exception of the maximum signal. A Gaussian fit is performed on the remaining data points. The FWHM and R\(^{2}\)-value of the fits are displayed in Fig. \ref{fig:IRF01_Experimental_time_resolution}b-c, respectively, with kinetic traces and their fits at selected probe wavelengths displayed in Fig. \ref{fig:IRF01_Experimental_time_resolution}d-i. The two methods provide similar measures of the CA width for wavelengths above 450 nm. Deviations occur for shorter wavelengths as the number of peaks and troughs increases. Overall, the half width of the total modulated interval is the most informative measure as it indicates what early time points can be included in a spectro-kinetic analysis of the TA data. From Fig. \ref{fig:IRF01_Experimental_time_resolution}b it is apparent that this is confined to \(\sim\)50 fs in the region 490--800 nm, less than 100 fs for 315--490 nm and less than 430 fs down to 200 nm.

The half width of the CA in the DUV is up to eight times wider than in the visible region. To rationalize this observation, we here discuss how the CA depends on the probe pulse's group velocity and group velocity dispersion (GVD) in more detail. To this end, we calculate the probe's wavelength dependent group delay (T\(_{\text{g}}\)), from the setup's optical pathlength and determine the accumulated T\(_{\text{g}}\) of the probe at the front face, center, and back face of the flow cell. Fig. \ref{fig:IRF02_Probe_chirp_analysis} displays the result in panel (d), with panels (a-c) plotting the underlying input parameters for all involved dispersive media. As the starting point for the calculation, Fig. \ref{fig:IRF02_Probe_chirp_analysis}a shows the refractive indices of all materials present between the exit of HCF3 and the back of the flow cell, as well as CaF\(_{2}\) as a commonly employed material for comparison. The refractive indices were obtained from ref. \cite{polyanskiy2024refractiveindex} (data from database: Air, Peck and Reeder, 1972; SiO\(_{2}\), Franta \textit{et al.}, 2016; H\(_{2}\)O, Daimon and Masumura, 2007, 20\(^{\circ}\)C; MgF\(_{2}\), Li, 1980, ordinary axis; CaF\(_{2}\), Li, 1980; accessed 03/22/2026), with the exception of helium for which the index of refraction at 5.9 bar was obtained from the modified Sellmeier equation in ref. \cite{ermolov2015supercontinuum}. From this data, the GVD is calculated in \ref{fig:IRF02_Probe_chirp_analysis}b. Although gasses generally have a much lower GVD, their contribution can become relevant for sufficiently long pathlengths. We also note here that out of the most common transparent materials with high transmission in the DUV, MgF\(_{2}\) presents the lowest GVD, which is an important advantage compared to CaF\(_{2}\) and SiO\(_{2}\). Fig. \ref{fig:IRF02_Probe_chirp_analysis}c then displays the accumulated T\(_{\text{g}}\) of the probe for the implemented TA setup up to the center of the flow cell, considering the optical path in each material separately. Note that the obtained group delay is plotted relative to the value at 510 nm to directly illustrate the walk-off between the probe and the pump pulses. Finally, Fig. \ref{fig:IRF02_Probe_chirp_analysis}d shows the sum of all T\(_{\text{g}}\) contributions at the front and back face of the flow cell and at its center, overlayed with the measured CA in H\(_{2}\)O. An excellent overall match between the calculated total T\(_{\text{g}}\) and the CA is observed, showing that the total T\(_{\text{g}}\) at the flow cell center follows the center of the CA. The T\(_{\text{g}}\) calculated at the flow cell's front and back face follows the curvature of the observed outermost XPM modulations.

The correspondence between the T\(_{\text{g}}\) at the flow cell's faces and the outermost signals of the CA in the DUV suggests that the temporal walk-off effect is the dominant contribution to the broadening of the CA below 380 nm. Besides the walk-off, the total curvature of T\(_{\text{g}}\) indicates that the chirp rate decreases in the DUV. While this causes subsequent broadening of the generated XPM, this contribution to the temporal width of the CA is observed to be much smaller than the contribution from the walk-off effect.

\paragraph{Effective response function}

Higher-order dispersion in the UV region of the probe means that the ERF cannot be approximated as a cross-correlation between the pump duration and the FTL of the probe \cite{polli2010effective}. Instead, upper limits to the ERF are established through the period of coherent oscillations and the 2PA contribution at the center of the CA. The solvent TA response displays coherent oscillations arising from the flow cell from 310 to 400 nm and 430 to 780 nm. A Fast Fourier Transform of the signal in both spectral regions resolves maximum Raman frequencies of 610 and 860 cm\(^{-1}\), respectively, corresponding to a period of \(\sim\)55 fs and \(\sim\)40 fs. Below 225 nm, the isolated 2PA from the solvent is a measure of the wavelength-dependent cross-correlation between the probe and the pump. The ERF is obtained as the FWHM of Gaussian fits of the 2PA signal and yields 160 fs and 95 fs at 200 nm and 220 nm, respectively.

\paragraph{Effective time resolution of the setup}

It is concluded that the temporal walk-off between the XPM signals from the flow cell's front and back surfaces is the dominant contribution to the width of the CA in the DUV. This can be observed at a probe wavelength of 200 nm (Fig. \ref{fig:IRF01_Experimental_time_resolution}a), where the temporal width of the 2PA signal is substantially narrower than the total CA width. This analysis demonstrates that the effective time resolution, here denoting the width of the CA, of the TA measurement not only depends on the durations of the employed pump and probe pulses, but also on their temporal walk-off along the propagation length of the sample. While this walk-off will have a significant impact on any TA measurement with pump and probe pulses of sufficiently short duration and spectral separation, it is particularly relevant in the reported setup: in the DUV below 300 nm the pump-probe walk-off with respect to the commonly employed visible pump pulses increases nonlinearly, due to the strong increase of the GVD of most employed optical materials. Therefore, to further improve the time resolution of the setup, reducing the walk-off accumulated by the probe throughout the interaction medium is critical. As suggested by the group delay contributions displayed in Fig. \ref{fig:IRF02_Probe_chirp_analysis}c, the flow cell windows contribute more than the solvent's sample volume. Indeed, previous TA setup implementations have shown that the use of a free liquid jet leads to substantially improved effective time resolution, reducing the temporal walk-off between pump and probe, while also removing the XPM contributions from any cell windows altogether \cite{godinez2026ultrafast}. Additionally, to obtain the highest possible time resolution, the entire pump-probe setup can be implemented inside an evacuated system, thereby minimizing the ERF  \cite{brahms2025decoupled}. However, we here choose to work with a sample cell-based flow system at atmosphere for ease of access and for the higher flexibility in solvent choice and sample volume afforded by a flow cell system.

\subsubsection{Preparation and characterization of [Fe(bpy)\(_{3}\)]Cl\(_{2}\) sample}

\begin{figure}[t]
\centering\includegraphics{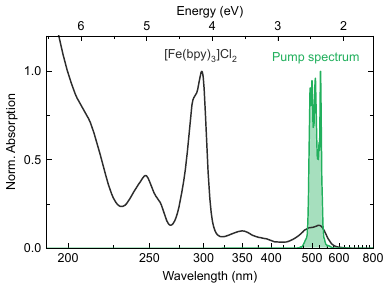}
\caption{Steady-state absorption of [Fe(bpy)\(_{3}\)]Cl\(_{2}\) in H\(_{2}\)O, normalized to the ligand centered transition at 298 nm. Also shown is the pump spectrum used in TA measurements.}
\label{fig:TA04_Steady_state_absorption}
\end{figure}

Tris(2,2'-bipyridine)iron(II) chloride ([Fe(bpy)\(_{3}\)]Cl\(_{2}\)) was synthesized following a modified literature procedure based on ref. \cite{field2016spectral}, and dissolved in MilliQ H\(_{2}\)O for measurements. Steady-state absorption spectra of solutions are measured in a 1 mm pathlength cuvette on a Shimadzu UV-2600 spectrometer. Fig. \ref{fig:TA04_Steady_state_absorption} shows the steady-state absorption spectrum of [Fe(bpy)\(_{3}\)]Cl\(_{2}\) as well as the pump spectrum used in TA, which overlaps with the metal-to-ligand charge transfer (MLCT) band of the complex.

\subsubsection{Conditions and procedures for TA measurements of [Fe(bpy)\(_{3}\)]Cl\(_{2}\)} \label{sec:SI_TA_methods}

\paragraph{Measurement conditions}

During TA measurements, the sample solution is circulated through a 100 \textmu m pathlength flow cell with 200 \textmu m thick thin-wall-aperture windows (Starna, 48/UTWA2/Q/0.1). The pump focal spot 1/e\(^{2}\) diameter is \(\sim\)240 \textmu m and the one of the probe varies between 50-80 \textmu m depending on the probe wavelength. The pump polarization direction is set at the magic angle (\(\sim 54.7^{\circ}\)) with respect to the probe. 

For the reported measurements, the OD of [Fe(bpy)\(_{3}\)]Cl\(_{2}\) in the 100 \textmu m pathlength flow cell is 0.33 at 200 nm, 0.33 at 298 nm and 0.043 at 521 nm, corresponding to a sample concentration of 0.56 mM. At each pump-probe time delay, 20'000 shots are detected and processed to calculate a TA spectrum. A separate B-matrix is determined from 10'000 shots at the beginning of each pump-probe delay scan and used to reference the first half of the following and the second half of the previous scan, except for the last scan that only uses the B-matrix prior to the scan. The longterm validity of a calculated B-matrix was tested by correcting later TA acquisitions, showing that the noise level only increases marginally within 1 hour (see Supplementary Fig. \ref{fig:Noise_SI_05_bmat}). For the [Fe(bpy)\(_{3}\)]Cl\(_{2}\) measurements, a single scan contains 451 (371) delay points for DUV (visible) TA, making for a total approximate scan time of 12 (14) minutes. A total of 20 scans are measured for DUV and visible TA and processed for averaging. For the presented TA data, the spectral regions are processed and averaged separately and then merged to form a broadband TA map.

\paragraph{Fluence dependence}

\begin{figure}[t!]
\centering\includegraphics[]{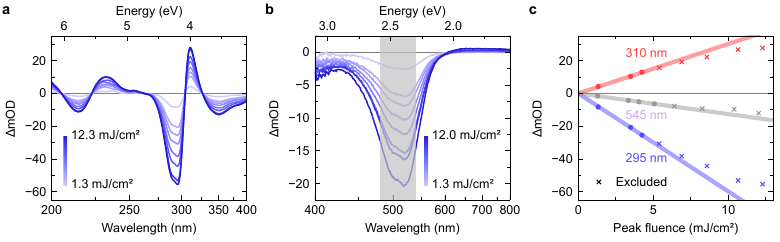}
\caption{Fluence dependence of [Fe(bpy)\(_{3}\)]Cl\(_{2}\) in H\(_{2}\)O prior to TA measurements, measured at a pump-probe delay of \(\sim\)10 ps. a,b) TA signal for various pump peak fluences for the two separate measurements in the regions 200-400 nm and 400-800 nm. c) Signal of selected TA features. Points within the linear regime are indicated with dots and used for a linear fit function.}
\label{fig:TA03_Fluence_dependence}
\end{figure}

Prior to TA measurements, the maximum pump peak fluence was determined for which the signal scales linearly. The peak fluence is calculated as \(F=8E/(\pi w^2)\), where \(E\) is the pulse energy and \(w\) the 1/e\(^{2}\) diameter of a Gaussian beam profile (\(w_{\text{major}}\sim w_{\text{minor}}\sim 240\) \textmu m). Fig. \ref{fig:TA03_Fluence_dependence} shows the TA measured in the DUV and visible regions, as well as the signal at their main features. Across both spectral regions, linearity is maintained until \(F\approx\) 4.1 mJ/cm\(^{2}\), corresponding to a pump energy of 0.92 \textmu J, which was hence used for measurements.

\subsubsection{[Fe(bpy)\(_{3}\)]Cl\(_{2}\) TA data and analysis}\label{sec:SI_GLA}

[Fe(bpy)\(_{3}\)]Cl\(_{2}\) in H\(_{2}\)O has been extensively studied by TA in different spectral regions and levels of time resolution, with pronounced excited-state signatures in the UV region below 400 nm offering direct insights into its electronic and structural dynamics \cite{consani2009vibrational,aubock2015sub,oppermann2022chiral}. These reports describe the excited-state dynamics following photo-excitation from the low-spin (LS) ground state through the metal-to-ligand charge-transfer (MLCT) band centered approximately at 520 nm. Intersystem crossing to a high-spin (HS) metal-centered state occurs on a sub-50 fs timescale \cite{aubock2015sub}, after which excess energy in the HS state is dissipated over the course of a few ps through vibrational relaxation, followed by ground-state recovery with a time constant of approximately 680 ps \cite{consani2009vibrational,aubock2015sub,oppermann2022chiral}.

The TA data of [Fe(bpy)\(_{3}\)]Cl\(_{2}\) in H\(_{2}\)O obtained with the present setup is displayed in Fig. \ref{fig:TA_Main} in the Main. This section presents the analysis of the data through a Global Lifetime Analysis (GLA) in the OPTIMUS program \cite{slavov2015implementation}. The fit is performed on the full time-zero-corrected map while excluding the region where the DUV and visible TA are merged. Additionally, data at pump-probe delays below 0.5 ps and at the pump wavelengths are excluded to avoid contributions from the CA and pump scatter, respectively. For both parallel and sequential fit models a minimum number of three exponential decays are required to reproduce all kinetic features of the measured TA data and removing one component leads to substantial deviations of the fits from the data.

\begin{figure}[t!]
\centering\includegraphics{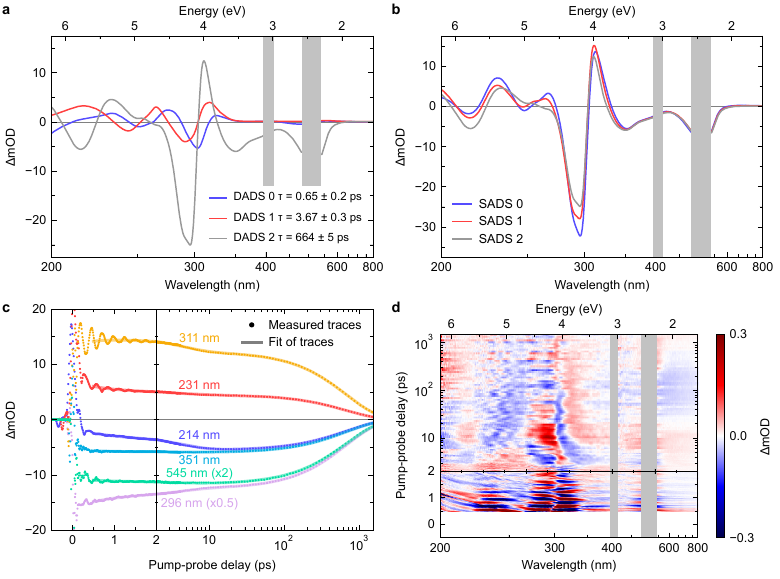}
\caption{Global lifetime analysis of [Fe(bpy)\(_{3}\)]Cl\(_{2}\) in H\(_{2}\)O. a) Decay-associated difference spectra with the corresponding lifetimes indicated in the legend. b) Species-associated difference spectra corresponding to the components in (a). c) Comparison between the data and GLA fit for the main features. d) Map of fit residuals.}
\label{fig:TA02_GLA_fits}
\end{figure}

Fig. \ref{fig:TA02_GLA_fits}a-b shows the Decay-Associated Difference Spectra (DADS) and Species-Associated Difference Spectra (SADS), respectively, obtained from a sequential fit model with three exponential decay components. A sequential fit model was used following previous reports \cite{oppermann2022chiral}, although a parallel fit model gives the same time constants within the respective error ranges. The obtained time constants are \(\tau_{1}\) = 0.65\(\pm\)0.20 ps, \(\tau_{2}\) = 3.7\(\pm\)0.3 ps and \(\tau_{3}\) = 664\(\pm\)5 ps. A comparison between the data and the GLA fit at the main features (Fig. \ref{fig:TA02_GLA_fits}c) and a map of the residuals (Fig. \ref{fig:TA02_GLA_fits}d) indicate that three lifetime components accurately describe the data. The time constants and spectral shapes of the DADS are in excellent agreement with previously recorded TA in the mid- and near-UV regions \cite{consani2009vibrational,aubock2015sub,oppermann2022chiral}. The values are assigned to vibrational cooling within the HS state (\(\tau_{1}\), \(\tau_{2}\)) \cite{aubock2015sub,oppermann2022chiral} and ground-state recovery (\(\tau_{3}\)) \cite{miller2020outer,oppermann2022chiral}. The GLA analysis indicates that the newly accessed region from 200 to 270 nm provides the same kinetic information as the data at longer wavelengths.

\subsubsection{Second order diffraction in grating spectrometer} \label{sec:SI_2nd_order_diffraction}
\begin{figure}[t!]
\centering\includegraphics{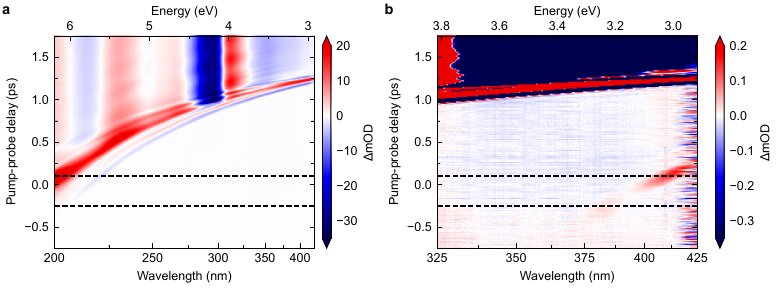}
\caption{Second-order diffraction signal in the TA spectrograph. a) TA map of [Fe(bpy)\(_{3}\)]Cl\(_{2}\) in the DUV before time-zero correction and prior to cropping the data to 400 nm. b) Zoom of (a) in the region of the second order diffraction. Note that the scale bar has reduced by 100x. The dashed lines indicate the second-order diffraction signal at 385 and 410 nm.}
\label{fig:TA05_Full_UV_map}
\end{figure}

One of the disadvantages of a grating spectrograph is that the second order diffraction of a wavelength overlaps with the first order diffraction of twice its value. As a result, a single measurement can only be performed over at most one spectral octave and separate measurements are required to probe the entire continuum out of HCF3. Fig. \ref{fig:TA05_Full_UV_map} displays the TA data of the DUV measurement extended on the red side to the full extent of the detector chip. There is a faint signal at \(\sim\)385 nm and \(\sim\)410 nm (note the scale bar reduced by x100) that is at double the wavelength of the CA in the DUV, suggesting that the probe spectrum extends to about 190 nm. The signal from the sample around 200 nm is 5 to 10 times weaker than that of the CA, hence its second-order diffraction is not observed. Therefore, it does not affect the first-order diffraction TA signal below 400 nm.

\clearpage

\putbib
\end{bibunit}

\end{document}